\pdfoutput=1
\documentclass[a4paper,11pt]{article}
\usepackage{amsmath,amssymb,amsfonts}
\usepackage[noconfig]{refstyle}
\usepackage{cite}
\usepackage[dvipsnames]{xcolor}
\usepackage[hyperfootnotes=false, linktocpage=true, colorlinks, citecolor=blue, linkcolor=blue, urlcolor=Maroon, filecolor=Maroon]{hyperref}
\usepackage{geometry,longtable}
\usepackage[small,bf]{caption}
\usepackage{tikz}
\usetikzlibrary{shapes,arrows}
\tikzstyle{block} = [rectangle, draw, text width=7em, text centered, rounded corners, minimum height=3em]

\usepackage[normalem]{ulem}

\let\eqref=\relax
\newref{eq}{name={},Name={Eq.~},names={eqs.~},Names={Eqs.~},rngtxt={-},refcmd=(\ref{#1})}
\newref{tab}{name={},Name={Table~},names={tables~},Names={Tables~}}
\newref{sec}{name={},Name={Section~},names={sections~},Names={Sections~}}
\newref{fig}{name={figure~},Name={Figure~},names={figures~},Names={Figures~}}
\numberwithin{equation}{section}

\newcommand{\be}{\begin{equation}}
\newcommand{\ee}{\end{equation}}

\def\fnote#1#2{\begingroup\def\thefootnote{#1}\footnote{#2}
     \addtocounter{footnote}{-1}\endgroup}
\DeclareMathOperator{\hook}{\lrcorner}

\usepackage{ulem}

\begin{document}

\vspace{1cm}

\title{\vspace{40pt}
       {\Large \bf Calabi-Yau Manifolds and $\boldsymbol{SU(3)}$ Structure}\\[2mm]
}

\author{
Magdalena Larfors${}^{1}$,
Andre Lukas${}^{2}$ and
Fabian Ruehle${}^{2}$
}
\date{}
\maketitle
\begin{center} {\small 
  ${}^1${\it Department of Physics and Astronomy, Uppsala University,\\  SE-751 20 Uppsala, Sweden}\\[4mm]
  ${}^2${\it Rudolf Peierls Centre for Theoretical Physics, University of Oxford,\\ Parks Road, Oxford OX1 3PU, UK}}\\

  \fnote{}{magdalena.larfors@physics.uu.se}
  \fnote{}{lukas@physics.ox.ac.uk} 
  \fnote{}{fabian.ruehle@physics.ox.ac.uk}
\end{center}

\begin{abstract}
\noindent
We show that non-trivial $SU(3)$ structures can be constructed on large classes of Calabi-Yau three-folds. Specifically, we focus on Calabi-Yau three-folds constructed as complete intersections in products of projective spaces, although we expect similar methods to apply to other constructions and also to Calabi-Yau four-folds. Among the wide range of possible $SU(3)$ structures we find Strominger-Hull systems, suitable for heterotic or type II string compactifications, on all complete intersection Calabi-Yau manifolds. These $SU(3)$ structures of Strominger-Hull type have a non-vanishing and non-closed three-form flux which needs to be supported by source terms in the associated Bianchi identity. We discuss the possibility of finding such source terms and present first steps towards their explicit construction. Provided suitable sources exist, our methods lead to Calabi-Yau compactifications of string theory with a non Ricci-flat, physical metric which can be written down explicitly and in analytic form.
\end{abstract}

\thispagestyle{empty}
\setcounter{page}{0}
\newpage

\tableofcontents
\newpage 

\section{Introduction}\seclabel{intro}
Calabi-Yau (CY) manifolds together with their Ricci-flat metrics lead to a large class of string solutions, which has been the main starting point for string compactifications and string model building to date. The construction of CY manifolds is relatively straightforward: apart from a complex and K\"ahler structure, only the vanishing of the first Chern class of the tangent bundle is required, a condition which is easily checked. Correspondingly, large classes of CY manifolds have been constructed, notably the traditional set of complete intersection CY three-folds in products of projective spaces (CICYs)~\cite{Candelas:1987kf,Green:1986ck,Hubsch:1992nu} and CY hypersurfaces in toric four-folds~\cite{Kreuzer:2000xy}. The existence of Ricci-flat metrics on CY manifolds is guaranteed by Yau's theorem \cite{Yau:1798aa}, although computing this metric is difficult and currently only possible with numerical methods~\cite{Donaldson:2005aa, Douglas:2006rr, Braun:2007sn, Douglas:2015aga}. Fortunately, some of the physics of CY compactifications, in particular the spectrum of massless particles and holomorphic quantities in the low-energy theory, can be extracted with methods of algebraic geometry without any recourse to the metric.  However, other crucial pieces of physics, such as the physical Yukawa couplings, do depend on the metric and are quite difficult to compute.

Another problematic aspect of CY compactifications is  moduli stabilisation. Successful stabilisation of moduli seems to require flux which leads into the realm of manifolds with $SU(3)$ structure. An $SU(3)$ structure on a six-dimensional manifold $X$ can be defined by a pair $(J,\Omega)$ of a two-form $J$ and a three-form $\Omega$, subject to certain constraints. As we will review, such structures are classified by five torsion classes $W_1,\ldots ,W_5$.  Manifolds with $SU(3)$ structure carry a globally defined spinor and, therefore, may preserve some supersymmetry in the context of string compactifications. Whether a given manifold with $SU(3)$ structure does indeed provide a solution to string theory depends on the pattern of torsion classes which have to match constraints arising from the chosen flux (as well as the dilaton solution and the geometry of the un-compactified space). The case of CY manifolds with a Ricci-flat metric corresponds to the special case of $SU(3)$ structure where all torsion classes vanish, $W_1=\cdots =W_5=0$. A closely related case is that of a CY manifold with a metric obtained by conformal re-scaling of the Ricci-flat metric, which leads to an $SU(3)$ structure characterised by $W_1=W_2=W_3=0$ and $3W_4=2W_5$.  In the following, we will refer to all other cases as non-trivial $SU(3)$ structures.

Although manifolds with non-trivial $SU(3)$ structures form a very large class,\footnote{Indeed, any six-dimensional, orientable spin manifold allows a reduction of the structure group to $SU(3)$ \cite{2005math......8428B}. However, this existence result does not determine the torsion of the $SU(3)$ structure.} {a limited set of} examples which fit into string theory have been constructed. The explicitly known geometries, suitable for heterotic string compactifications, consist of either homogeneous examples~\cite{Fernandez:2008wa,Grantcharov:2009qv,Fei:2014aca,Otal:2016bgn} or torus fibrations over certain four-dimensional base spaces~\cite{Goldstein:2002pg,Fu:2006vj,Fei:2017ctw}. An interesting solution-generating method is also provided in Ref.~\cite{Martelli:2010jx}. For type II compactifications, example geometries have been constructed on twistor spaces~\cite{Nilsson:1984bj,SOROKIN1985301,Tomasiello:2007eq},  {on more generic cosets~\cite{Koerber:2008rx}, solvmanifolds~\cite{Andriot:2015sia} and on toric varieties~\cite{Larfors:2010wb,Larfors:2011zz,Terrisse:2017tbb}. While these geometries have led to interesting examples of string vacua, their construction is somewhat tedious.} The problem is that there is no analogue of Yau's theorem for the types of non-trivial $SU(3)$ structures required by string theory. This means that $SU(3)$ structures have to be constructed explicitly, for example by constructing the forms $(J,\Omega)$. Only after this, often laborious, task is completed is it possibly to decide whether the structure is compatible with string theory. Frequently, the answer turns out to be ``no" and one has to start over in this ``unguided" search for suitable $SU(3)$ structures. On the upside, once an $SU(3)$ structure relevant to string theory has been found explicitly, the associated metric can be computed from the forms $(J,\Omega)$ and is therefore readily available for computing quantities in the associated effective theory, such as physical Yukawa couplings.

The above discussion suggests that string compactification involves a choice between two options, both of which have considerable downsides. Calabi-Yau compactifications have the benefit of relying on large classes of available spaces whose algebraic properties are well explored. However, the differential geometry of CY manifolds, which is ultimately required in a string theory context, is hugely inaccessible due to the difficulties in computing the Ricci-flat metric. On the other hand, compactifications on manifolds with a non-trivial $SU(3)$ structure, offer some hope of a more accessible differential geometry. However, very few such spaces relevant to string theory are known.\footnote{We note that the conformal CY case {\it e.g.} \cite{Gukov:1999ya}, does not offer a more accessible differential geometry since it is based on the unknown Ricci-flat metric. The same applies to the heterotic flux vacua that are constructed as deformations of CY solutions, {\it e.g.} \cite{Witten:1985bz,Witten:1986kg}, and more recently \cite{li2005,Andreas:2010cv,Andreas:2010qh}. This is why we have excluded these cases from what we have called non-trivial $SU(3)$ structures.}

In this paper, we explore whether the advantages of CY compactifications and those of compactifications on manifolds with $SU(3)$ structure can be combined in a new approach. Specifically, we would like to ask and answer the following two questions:
\begin{itemize}
 \item Can CY manifolds carry non-trivial  $SU(3)$ structures with explicit metrics?
 \item If so, are these non-trivial, explicit $SU(3)$ structures on CY manifolds relevant for string compactification?
\end{itemize} 
We will answer the first of these questions with a resounding ``yes" and the second one with a somewhat tentative ``yes".

The plan of the paper is as follows. In the next section, we start with a brief review of $SU(3)$ structure, in order to prepare the ground and set the notation. Section~\secref{quintic} studies the possibility of non-trivial $SU(3)$ structures on the quintic, as a warm-up exercise. In Section~\secref{cicys}, these results will be generalised to all CICY manifolds. Further explicit examples are presented in Section~\secref{examples} and in Section~\secref{BI} we discuss the requirement of satisfying the Bianchi identity for the flux. We conclude in Section~\secref{conclusion}. Some technical results relevant to the discussion in the main text are collected in Appendix~\secref{appA}.

\section{\texorpdfstring{$\boldsymbol{SU(3)}$}{SU(3)} structure}\seclabel{SU3}
In this section, we review a number of well-known facts about $SU(3)$ structures (see, for example, Refs.~\cite{2000math.....10054H,2002math......2282C,LopesCardoso:2002vpf} for more detailed accounts), in order to set the scene and fix our notation. We begin with the mathematical background and then move on to some aspects of $SU(3)$ structures relevant in the context of string theory.

\subsection[Definition and properties of \texorpdfstring{$SU(3)$}{SU(3)} structure]{Definition and properties of \texorpdfstring{$\boldsymbol{SU(3)}$}{SU(3)} structure}
An $SU(3)$ structure of a six-dimensional manifold $X$ is defined as a sub-bundle of the frame bundle which has structure group $SU(3)$. This means there are local frames $\epsilon^A$, where $A=1,\ldots ,6$, of the (co-)tangent bundle which patch together with $SU(3)$ transition functions. More explicitly, if we introduce a frame $e^a$, where $a=1,2,3$, together with the complex conjugates $\bar{e}^a$ of the (complexified) co-tangent bundle by
\begin{equation}
 e^1=\epsilon^1+i\epsilon^2\;,\quad e^2=\epsilon^3+i\epsilon^4\;,\quad e^3=\epsilon^5+i\epsilon^6\; ,
\end{equation}
then the presence of an $SU(3)$ structure corresponds to the frame transformations $e^a\rightarrow {U^a}_be^b$ with $U\in SU(3)$. It is then immediately clear that the two forms
\begin{equation}
 J=\frac{i}{2}\sum_a e^a\wedge \bar{e}^a\;,\qquad \Omega= e^1\wedge e^2\wedge e^3\; . \eqlabel{JOlocal}
\end{equation} 
are $SU(3)$ invariant and, hence, globally well-defined on $X$. The metric associated to this $SU(3)$ structure is given by
\begin{equation}
 g=\sum_ae^a\otimes \bar{e}^a\; .
\end{equation}
In terms of the real six-bein $\epsilon^A$ the above forms can also be written as
\begin{equation}
\begin{array}{l@{~}l@{~}lll@{~}l@{~}l}
 J&=&\epsilon^{12}+\epsilon^{34}+\epsilon^{56}\;,&& \Omega&=&\Omega_++i\Omega_-\\
  \Omega_+&=&\epsilon^{135}-\epsilon^{146}-\epsilon^{236}-\epsilon^{245}\;,&&
 \Omega_-&=&\epsilon^{136}+\epsilon^{145}+\epsilon^{235}-\epsilon^{246}\; ,
\end{array} 
\end{equation} 
where $\epsilon^{12}$ is a short-hand notation for $\epsilon^1\wedge\epsilon^2$, and similar for the other expressions of this type. 

Alternatively, an $SU(3)$ structure on the six-dimensional manifold $X$ can be specified by a pair $(J,\Omega)$ of a globally-defined real two-form $J$ (which has to be positive everywhere on $X$) and a globally-defined three-form $\Omega$ on $X$ satisfying
\begin{equation}
 J\wedge J\wedge J=\frac{3i}{4}\,\Omega\wedge\bar{\Omega}\;,\qquad J\wedge\Omega=0\; . \eqlabel{JOdef}
\end{equation}
Note that the (local) expressions~\eqref{JOlocal} for $J$ and $\Omega$ do indeed satisfy those equations. Conversely, forms $(J,\Omega)$ satisfying~\eqref{JOdef} can always be written locally as in~\eqref{JOlocal}.

The torsion classes can be read off from the exterior derivatives of $J$ and $\Omega$, which can be cast into the form
\begin{equation}
 dJ=\frac{3i}{4}\left(\bar{W}_1\Omega+W_1\bar{\Omega}\right)+W_4\wedge J+W_3\;,\quad
 d\Omega=W_1J\wedge J+W_2\wedge J+W_5\wedge\Omega\; , \eqlabel{dJdO}
\end{equation}
where we decompose $W_1=W_1^++iW_1^-$ and $W_2=W_2^++iW_2^-$. The torsion classes $W_1$, $W_4$ and $W_5$ can be computed from $dJ$ and $d\Omega$ via
\begin{equation}
 W_1=-\frac{i}{6}\Omega \hook dJ=\frac{1}{12}J^2 \hook d\Omega\;,\quad
 W_4=\frac{1}{2}J\hook dJ\;,\quad
 W_5=-\frac{1}{2}\Omega_+\hook d\Omega_+\; ,
\end{equation} 
where the contraction of forms is normalised such that, for example, $\epsilon^{12}\hook \epsilon^{1234} =  \epsilon^{34}$. The remaining torsion classes are constrained by
\begin{equation}
 J\hook W_2=0\;,\quad J\hook W_3=0\;,\quad \Omega\hook W_3=0\; .
\end{equation} 
Altogether, this implies the $SU(3)$ representations for the torsion classes as given in Table~\tabref{torsionSU3}.
\begin{table}[t]
\begin{center}
 \begin{tabular}{|c|c|}\hline
  torsion class&$SU(3)$ representation\\\hline\hline
  $W_1$&${\bf 1}\oplus{\bf 1}$\\\hline
  $W_2$&${\bf 8}\oplus{\bf 8}$\\\hline
  $W_3$&${\bf 6}\oplus\bar{\bf 6}$\\\hline
  $W_4$&${\bf 3}\oplus\bar{\bf 3}$\\\hline
  $W_5$&${\bf 3}\oplus\bar{\bf 3}$\\\hline
\end{tabular}
\caption{Torsion classes for $SU(3)$ structures and their associated $SU(3)$ representations.}\tablabel{torsionSU3}
\end{center}
\end{table}  
For illustration and later reference, Table~\tabref{ta:manifolds} lists a number of mathematical properties of six-dimensional manifolds and the associated vanishing pattern of the torsion classes.
\begin{table}[t]
\begin{center}
\begin{tabular}{|c|c|}
\hline
{\bf Property} & {\bf Vanishing torsion class}\\
\hline
Complex & $W_1=W_2=0$ \\ 
\hline
Half-flat & $W_1^-= W_2^-=W_4=W_5=0$ \\
\hline
Special Hermitean & $W_1=W_2=W_4=W_5=0$ \\ 
\hline
Nearly K\"ahler & $W_2=W_3=W_4=W_5=0$ \\
\hline
Almost K\"ahler & $W_1=W_3=W_4=W_5=0$ \\
\hline
K\"ahler & $ W_1=W_2=W_3=W_4=0$ \\
\hline
Ricci-flat & $ W_1=W_2=W_3=W_4=W_5=0$ \\
\hline
conformal Ricci-flat & $ W_1=W_2=W_3= 3 W_4-2 W_5=0$ \\
\hline
\end{tabular}
\caption{Mathematical properties and associated pattern of torsion classes, taken from Ref.~\cite{Grana:2005jc}. Superscripts $\pm$ indicate, respectively, the real or imaginary part of the torsion class.}\tablabel{ta:manifolds}
\end{center}
\end{table}

\subsection[\texorpdfstring{$SU(3)$}{SU(3)} structure in string theory]{\texorpdfstring{$\boldsymbol{SU(3)}$}{SU(3)} structure in string theory}
In string theory, one is interested in 10-dimensional spacetimes $\hat{M}$ of the form $\hat{M}=X\times M$, where $X$ is a six-dimensional compact manifold and $M$ is the four-dimensional non-compact spacetime. The metric on $\hat{M}$ has product or warped product structure, with the metric on $X$ induced from an $SU(3)$ structure on $X$. This $SU(3)$ structure also implies the existence of a globally defined spinor, as is required to preserve a minimal amount of supersymmetry, a feature usually desirable in string theory.
 The required type of $SU(3)$ structure depends on the choice of flux, the dilaton profile, and the non-compact space $M$ and its metric. The latter is frequently chosen to be Minkowski space, AdS or dS with the maximally symmetric metric. To give a flavour of the relevant types of $SU(3)$ structures, we list various supersymmetric string compactifications  in Table~\tabref{tab:su3N1}, together with the required pattern of torsion classes on $X$. The interested reader may find more detailed accounts on these $SU(3)$ structure string vacua, and their various generalisations, in \cite{Hull:1986kz,Strominger:1986uh,Ivanov:2000fg,Giddings:2001yu,LopesCardoso:2002vpf,Becker:2003yv,Gauntlett:2003cy,Behrndt:2004km,Grana:2004bg,Lust:2004ig,Behrndt:2005bv,Larfors:2013zva} and the reviews \cite{Grana:2005jc,Blumenhagen:2006ci,Denef:2008wq}.
\begin{table}[t]
\begin{center}
\begin{tabular}{| l | l | l | l |}
\hline
\bf{4D geometry}\!\! & \bf{String vacuum} & \!\!$\begin{array}{l}\textbf{Non-vanishing}\\\textbf{torsion}\end{array}$\!\! & \!\!\bf{$\boldsymbol{SU(3)}$ type}\!\!  \\
\hline
\hline
${\cal N}=1$ Mkw & Heterotic, Type II ($H_3$, no~RR flux) & $W_3, W_4 = d \phi, W_5= 2W_4$& Complex\\
 \hline
${\cal N}=1$ Mkw & Type IIB ($H_3, F_3, F_5, O_3/O_7$) & $3W_4=2W_5$& Conf. CY\\
 & Type IIB/F-theory ($H_3, F_3, F_5, O_3/O_7$) & $W_4=W_5$& Complex\\
 & Type IIB ($F_3, O_5/O_9$) & $W_3, W_4 = d \phi, W_5= 2W_4$& Complex\\
\hline
${\cal N}=1$ Mkw & Type IIA ($F_2, F_4, O_6$) & $W_{2}, 3W_5 = d\phi$ & Symplectic \\
\hline
 ${\cal N}=1$ AdS & Type IIA   ($H_3, F_{\rm even}$) & $W_{1}^+, W_2^+, d W_2^+ \propto \Omega^+$ & Half-flat\\
 \hline
\end{tabular}
\caption{An overview of $SU(3)$ vacua of string theory where $H_3$ denotes the NS flux, $F_p$ the RR $p$-form flux, $O_p$ the orientifold planes and $\phi$ the dilaton. The third column lists necessary SUSY constraints on the torsion (necessary and sufficient for the type IIA AdS vacua). The table summarises information from Refs.~\cite{Grana:2005jc,Larfors:2013zva}.} \tablabel{tab:su3N1}
\end{center}
\end{table}

Of particular importance for the rest of the paper will be the Strominger-Hull system~\cite{Hull:1986kz,Strominger:1986uh,LopesCardoso:2002vpf} which is characterised by a pattern of torsion classes given by
\begin{equation}
W_1=W_2=0\;,\quad W_5=2W_4\;,\quad W_3\mbox{ arbitrary}\; . \eqlabel{TSH}
\end{equation}
From Table~\tabref{ta:manifolds}, this means the manifold $X$ has a complex structure but is, in general, not K\"ahler (they are, however, conformally balanced). The Strominger-Hull system can provide string solutions both in the case of the heterotic and the type II string. In either case, the dilaton $\phi$ is specified by
\begin{equation}
  W_4=d\phi\; , \eqlabel{phiSH}
\end{equation}
with the string coupling $g_S=e^\phi$. 

For the heterotic case, the NS three-form field strength $H$ can be expressed in terms of the torsion classes as
\begin{equation}
*H = e^{2\phi} d(e^{-2\phi} J) = - W_4 \wedge J + W_3 ~\iff~ H =  i (\partial - \bar{\partial}) J =2\, {\rm Im}\, (W_4^{(0,1)}\wedge J + W_3^{(1,2)} )  \; , \eqlabel{HSH}
\end{equation} 
where the superscripts denote the component of the torsion class with the indicated number of holomorphic and anti-holomorphic indices. Heterotic compactifications also involve a vector bundle $V\rightarrow X$ with connection $A$ and associated field strength $F$. In order for the gauge bundle to preserve supersymmetry, the connection has to satisfy the conditions
\begin{equation}
 F \wedge \Omega = 0\;,\quad F \wedge J \wedge J = 0\; . \eqlabel{Fsusy}
\end{equation} 
In addition, the field strength associated with this gauge connection is related to the curvature $R^-$ of the Hull connection $\nabla^-$ on $X$ by the Bianchi identity
\begin{equation}
d H=\tfrac{\alpha'}{4}\left({\rm{tr}}(F\wedge F)-{\rm{tr}}( R^- \wedge R^-)\right) \approx \tfrac{\alpha'}{4}\left({\rm tr}(F\wedge F)-{\rm tr}( R \wedge R )\right)\;. \eqlabel{eq:BI}
\end{equation}
Here, {$R^-$ is the curvature of the connection $\nabla^-$, whose Christoffel symbols are given by $\Gamma^{-m}_{np} = \Gamma^{m}_{np}-\frac{1}{2}H^m{}_{np}$}
 {(where $\Gamma^{m}_{np}$ are the symbols of the Levi--Civita metric associated to the $SU(3)$ structure)}, and the last approximation retains only leading order terms in the $\alpha'$ expansion.

{It should be noted that the $\alpha'$ corrections to the Bianchi identity \eqref{eq:BI} are required for solutions with non-vanishing $H$ flux on compact manifolds \cite{Gauntlett:2002sc}. Furthermore, while the constraints \eqref{TSH}-\eqref{eq:BI} can be shown to be necessary for a heterotic  ${\cal N}=1$ Minkowski solution, they do not directly imply the equations of motion}. As was shown in {Ref.~\cite{Ivanov:2000fg} (see also \cite{Martelli:2010jx,delaOssa:2014cia})}  one needs to require in addition that $\nabla^-$ is an $SU(3)$ instanton,
\begin{equation}
 R^- \wedge \Omega = 0\;,\quad R^- \wedge J \wedge J = 0\; . \eqlabel{eq:integrab}
\end{equation} 
To see that this condition is satisfied, we first note that the vanishing of the gravitino variation {$\delta \psi_m = \nabla^+_m \eta$} {implies} that $\nabla^+$ is an instanton. Second, it is straightforward to show that 
\begin{equation}
 R^+_{mnpq}-R^-_{pqmn} = \frac{1}{2} d H_{mnpq}\; , \eqlabel{eq:integrab2}
\end{equation} 
where the right hand side is $\cal{O}(\alpha')$ by the Bianchi identity \eqref{eq:BI}. { Hence,} $\nabla^-$ is {also} an $SU(3)$ instanton, but only up to first order $\alpha'$ corrections. However, these corrections appear at the same order {in $\alpha'$} as {other} terms that {have already been} neglected in the equations of motion, and should therefore be discarded {as well}. We refer the reader to Appendix A of \cite{delaOssa:2014cia} for a recent thorough discussion of this issue.

For type II string theory, $H$ can play the role of the NS three-form (as in the heterotic case), or it can be interpreted as the RR three-form. In the former case, the relations~\eqref{phiSH} and \eqref{HSH} remain valid, while the Bianchi identity for $H$ reads
\begin{equation}
 dH=\mbox{NS 5-brane sources}\; .
\end{equation}
If $H$ is identified with the RR three-form in type IIB string theory, the relation between the torsion classes and the flux is modified to
\begin{equation}
* H = e^{\phi}d (e^{-2\phi} J) = e^{-\phi} (-W_4 \wedge J + W_3) \;,
\end{equation} 
while the Bianchi identity now reads
\begin{equation}
 dH=\mbox{RR 5-brane sources}\; .
\end{equation} 
For much of the following discussion, we will be focusing on the pattern of torsion classes~\eqref{TSH} for the Strominger-Hull system and only return to the task of satisfying the Bianchi identity in Section~\secref{BI}.

\section{A warm-up example: the quintic}\seclabel{quintic}
In this section, we discuss possible non-trivial $SU(3)$ structures on the quintic CY, defined as the anti-canonical hypersurface in the ambient space ${\cal A}=\mathbb{P}^4$. This is a warm-up example for the next section, where this discussion will be generalised to all CICY manifolds. We begin with some general background and notation for the projective space $\mathbb{P}^4$.

\subsection{Basics}
On $\mathbb{P}^4$ we introduce homogeneous coordinates $x_A$, where $A=0,\ldots ,4$, and we define the standard patches $U_A=\{[x_0:\cdots :x_4]\,|\,x_A\neq 0\}$. We will frequently be working on the patch $U_0$ where we denote the affine coordinates by $z_a=x_a/x_0$, where $a=1,\ldots ,4$. Two useful quantities, which will appear throughout, are
\begin{equation}
 \sigma=\sum_{A=0}^4|x_A|^2\;,\qquad \kappa=1+\sum_{a=1}^4|z_a|^2\; ,
\end{equation}
where the first is the homogeneous version and the second its affine counterpart. In terms of these quantities, the Fubini-Study K\"ahler form ${\cal J}$ can be written as
\begin{equation}
 {\cal J}=\frac{i}{2\pi}\partial\bar{\partial}\ln\kappa =\frac{i}{2\pi}\sum_{a,b=1}^4\left[\frac{|z_a|^2}{\kappa}\delta_{ab}-\frac{|z_a|^2|z_b|^2}{\kappa^2}\right]\frac{dz_a}{z_a}\wedge\frac{d\bar{z}_b}{\bar{z}_b}\; . \eqlabel{FSP4}
\end{equation} 
The normalisation is chosen such that $\int_{\mathbb{P}^4}{\cal J}^4=1$.

A quintic $X\subset\mathbb{P}^4$ is defined as the zero locus of a polynomial $P=P(x)$ which is homogeneous of degree five in the coordinates $x_A$. The affine version of this polynomial is denoted by $p=p(z)$ and it is related to its homogeneous counterpart via
\begin{equation}
 p({z})=P(1,z)\;,\qquad P(x)=x_0^5\,p\left(\frac{x_1}{x_0},\ldots ,\frac{x_4}{x_0}\right)\; .
\end{equation} 
The above K\"ahler form ${\cal J}$ can be restricted to the quintic, which leads to a K\"ahler form
\begin{equation}
 J_0={\cal J}|_X \eqlabel{quinticJ}
\end{equation}
on $X$.  In practice, this restriction can be carried out by solving the defining equation $p=0$ for, say, $z_4$ in terms of the remaining three coordinates $z_\alpha$, $\alpha=1,2,3$, and by replacing the differential $dz_4$ with
\begin{equation}
 dz_4=-\sum_{\alpha=1}^3\frac{p_{,\alpha}}{p_{,4}}dz_\alpha\;,\qquad p_{,i}=\frac{\partial p}{\partial z_i}\; .
\end{equation}
Besides $J_0$, another standard differential form on $X$ is the $(3,0)$-form $\Omega_0$ which, on the patch $U_0$, can be explicitly written as~\cite{Witten:1985xc, Strominger:1985it}
\begin{equation}
 \Omega_0=\frac{dz_1\wedge dz_2\wedge dz_3}{p_{,4}} \eqlabel{quinticOmega}\; .
\end{equation}
We can ask if the above forms $(J_0,\Omega_0)$ already define an $SU(3)$ structure on $X$. Given the index structure of both forms we clearly have
\begin{equation}
 J_0\wedge \Omega_0=0\; , \eqlabel{JOquintic}
\end{equation}
so that the second condition~\eqref{JOdef} for an $SU(3)$ structure is satisfied. In order to check the first condition, we carry out an explicit calculation, using Eqs.~\eqref{FSP4}, \eqref{quinticJ} and \eqref{quinticOmega}, which leads to 
\begin{equation}
 J_0\wedge J_0\wedge J_0=\frac{3i}{4}{\cal F}\;\Omega_0\wedge\bar{\Omega}_0\; . \eqlabel{relquintic}
\end{equation} 
The function ${\cal F}$ on $X$ reads in homogeneous and affine form
\begin{equation}
 {\cal F}=\frac{1}{\pi^3\sigma^4}|\nabla P|^2=\frac{1}{\pi^3\kappa^4}\left[\sum_{a=1}^4|p_{,a}|^2+\sum_{a,b=1}^4z_a\bar{z}_bp_{,a}\bar{p}_{,b}\right]\; , \eqlabel{Fquintic}
\end{equation} 
respectively, where $\nabla P$ denotes the gradient of $P$ in terms of homogeneous coordinates. Since this function is non-trivial, the first conditions~\eqref{JOdef} is not satisfied and the pair $(J_0,\Omega_0)$ does not define an $SU(3)$ structure. 

We note from Eq.~\eqref{Fquintic} that ${\cal F}$ is well-defined on $\mathbb{P}^4$ since it is homogeneous of degree zero in $x_A$ and $\bar{x}_A$ and it is non-singular since $\sigma$ does not vanish on $\mathbb{P}^4$. Moreover, the quintic $X$ (defined by $P=0$) is smooth precisely if $\nabla P\neq 0$ everywhere on $X$, so ${\cal F}$ is a strictly positive function.

\subsection[\texorpdfstring{$SU(3)$}{SU(3)} structures on the quintic]{\texorpdfstring{$\boldsymbol{SU(3)}$}{SU(3)} structures on the quintic}
As we have seen, the above forms $(J_0,\Omega_0)$ do not define an $SU(3)$ structure due to the appearance of the non-trivial function ${\cal F}$ in Eq.~\eqref{relquintic}. However, this problem can be easily fixed by a conformal re-scaling. Indeed, by virtue of Eqs.~\eqref{JOquintic} and \eqref{relquintic}, the re-scaled forms
\begin{equation}
 J={\cal F}^kJ_0\;,\qquad \Omega={\cal F}^{\frac{3k+1}{2}}\Omega_0\; ,
\end{equation}
satisfy the relations~\eqref{JOdef} and, hence, define an $SU(3)$ structure for any real number $k$. Note, in particular, that ${\cal F}$ is strictly positive for a smooth quintic and, hence, $J$ is a positive form, as required.

What are the torsion classes associated to this $SU(3)$ structure? Using $dJ_0=d\Omega_0=0$, the exterior derivatives are easily computed as
\begin{equation}
 dJ=k\,d(\ln {\cal F})\wedge J\;,\qquad d\Omega=\frac{3k+1}{2}\,d(\ln {\cal F})\wedge\Omega\; .
\end{equation}
A comparison with the general equations~\eqref{dJdO} for these derivates shows that the torsion classes are given by
\begin{equation}
 W_1=W_2=W_3=0\;,\quad W_4=k\,d(\ln {\cal F})\;,\quad W_5=\frac{3k+1}{2}\,d(\ln {\cal F})\; . \eqlabel{structquintic}
\end{equation}  
Since $W_1$ and $W_2$ vanish, we know from Table~\tabref{tab:su3N1} that  we have an associated complex structure on $X$. If we further specialise to $k=1$, we find
\begin{equation}
 W_1=W_2=W_3=0\;,\quad W_5=2\,W_4=2\,d(\ln {\cal F})\; ,
\end{equation} 
which defines a Strominger-Hull system with $W_3=0$, as can be seen by comparing the expression with Eq.~\eqref{TSH}. The dilaton is fixed by
\begin{equation}
 d\phi=d(\ln {\cal F})\; ,
\end{equation}
so that the string coupling $g_S={\rm const}\times {\cal F}$ can be kept perturbative everywhere on $X$, for a suitable choice of the integration constant. If $H$ is interpreted as an NS flux, it is explicitly given by
\begin{equation}
 H=i(\partial-\bar{\partial})J=i{\cal F}^{-1}(\partial {\cal F}-\bar{\partial}{\cal F})\wedge J\; .
\end{equation} 

In conclusion, starting from the Fubini-Study metric on $\mathbb{P}^4$ and the standard $(3,0)$ form on the quintic, we can construct a family of $SU(3)$ structures, parametrised by a real number $k$, on every smooth quintic. For a special choice, $k=1$,  this $SU(3)$ structure is of the Strominger-Hull form with $W_3=0$. The dilaton varies non-trivially, but can be kept in the perturbative range and we have non-zero NS flux. 

We note that these $SU(3)$ structures do not corresponds to the conformally Ricci-flat case and are, hence, non-trivial in the sense defined earlier. Indeed, the last row in Table~\tabref{ta:manifolds} shows that a conformally Ricci-flat $SU(3)$ structure is characterised by $3W_4-2W_5=0$, while, from Eq.~\eqref{structquintic}, our $SU(3)$ structures satisfy
\begin{equation}
 3W_4-2W_5=-d(\ln {\cal F})\; .
\end{equation} 
Our next step will be to generalise this discussion to all CICY manifolds.

\section{CICYs and \texorpdfstring{$\boldsymbol{SU(3)}$}{SU(3)} structure}\seclabel{cicys}
We will now discuss complete intersection CY manifolds (CICYs) in the ambient space ${\cal A}=\mathbb{P}^{n_1}\times\cdots\times \mathbb{P}^{n_m}$. We first review some standard results and notation and then construct $SU(3)$-structures on these manifolds, generalising the approach we have taken for the quintic.

\subsection{Basics}
As mentioned above, the ambient space is given by a product ${\cal A}=\mathbb{P}^{n_1}\times\cdots\times \mathbb{P}^{n_m}$ of $m$ projective factors with dimensions $n_i$, where $i=1,\ldots ,m$, and total dimension $d=\sum_{i=1}^mn_i$. Homogeneous coordinates for each projective factor are denoted by ${\bf x}_i=(x_{iA})$, where $A=0,1\ldots ,n_i$. Their affine counterparts in the patch $x_{i0}\neq 0$ are called ${\bf z}_i=(z_{ia})$ with $z_{ia}=x_{ia}/x_{i0}$ and $a=1,\ldots ,n_i$. We will frequently work in the patch $U_0$ of ${\cal A}$ where all $x_{i0}\neq 0$, using the coordinates $({\bf z}_1,\ldots,{\bf z}_m)$. In analogy with the quintic case, we define the quantities
\begin{equation}
 \sigma_i=\sum_{A=0}^{n_i}|x_{iA}|^2\;,\qquad \kappa_i=1+\sum_{a=1}^{n_i}|z_{ia}|^2\; , \eqlabel{skdef}
\end{equation}
which can be used to write down the Fubini-Study K\"ahler forms ${\cal J}_i$ for each projective factor. In affine coordinates ${\bf z}_i$ they are explicitly given by
\begin{equation}
 {\cal J}_i=\frac{i}{2\pi}\partial\bar{\partial}\ln\kappa_i =\frac{i}{2\pi}\sum_{a,b=1}^{n_i}\left[\frac{|z_{ia}|^2}{\kappa_i}\delta_{ab}-\frac{|z_{ia}|^2|z_{ib}|^2}{\kappa_i^2}\right]\frac{dz_{ia}}{z_{ia}}\wedge\frac{d\bar{z}_{ib}}{\bar{z}_{ib}}\; . \eqlabel{FSgen}
\end{equation} 

The CICY three-fold $X$ is defined as the common zero locus of $K$ polynomials $P_u=P_u({\bf x_1},\!\ldots\! ,{\bf x}_m)$, $u=1,\ldots ,K$. They are homogeneous with multi-degree ${\bf q}_u=(q_u^i)$, where $q_u^i$ is the degree of homogeneity of $P_u$ in the coordinates ${\bf x}_i$ of the $i^{\rm th}$ projective factor. The polynomials are related to their affine counterparts $p_u=p_u({\bf z}_1,\ldots ,{\bf z}_m)$ on the patch $U_0$ by
\begin{equation}\left.
\begin{array}{ccc}
 p_u({\bf z}_1,\ldots ,{\bf z}_m)&=&P_u(1,{\bf z}_1,\ldots ,1,{\bf z}_m)\\
 P_u({\bf x}_1,\ldots ,{\bf x}_m)&=&\mathfrak{s}_u\, p_u\left(\frac{x_{1\,a}}{x_{1\,0}},\ldots ,\frac{x_{m\,a}}{x_{m\,0}}\right)
\end{array} \right\}\qquad
 \mathfrak{s}_u=\prod_{i=1}^mx_{i0}^{q_u^i}\; .
 \eqlabel{pAndP}
\end{equation} 

The information about the multi-degrees of the defining polynomials, together with the dimensions of the projective ambient space factors, is often summarised by the configuration matrix
\begin{equation}
 X\sim\left[\begin{array}{c|ccc}\mathbb{P}^{n_1}&q_1^1&\cdots&q_K^1\\\vdots&\vdots&&\vdots\\\mathbb{P}^{n_m}&q_1^m&\cdots&q_K^m\end{array}\right]^{h^{1,1},h^{2,1}}_{\eta}\; ,
\end{equation} 
where the two non-trivial Hodge numbers $h^{1,1}$ and $h^{2,1}$ are attached as superscripts and the Euler number $\eta=2(h^{1,1}-h^{2,1})$ as a subscript. The Calabi-Yau condition, $c_1(TX)=0$, simply translates into the conditions $\sum_{u=1}^Kq_u^i=n_i+1$, for $i=1,\ldots ,m$, on the degrees. Using this notation, the quintic in $\mathbb{P}^4$ discussed in the previous section is described by the configuration $[\mathbb{P}^4\;|\;5]^{\mbox{\tiny$1,101$}}_{\mbox{\tiny $-200$}}$.

There is an infinite number of CICY configuration matrices of the above type but it turns out that different configurations can correspond to the same topological class of Calabi-Yau manifolds. Taking this identification into account, the number of topological types of CICY three-folds becomes finite and the classification of Ref.~\cite{Candelas:1987kf, Green:1986ck} leads to 7890 topological types.\footnote{The full list, including supplementary information, can be downloaded from~\url{http://www-thphys.physics.ox.ac.uk/projects/CalabiYau/cicylist/}} This list provides one representative configuration matrix for each topological type and we will use some examples from this list in the next section.
For now the discussion will be carried out in terms of a general configuration matrix and, hence, applies to all CICY manifolds and all configuration matrices.

There are a number of obvious differential forms which can be defined on $X$. First of all we can restrict the Fubini-Study K\"ahler forms~\eqref{FSgen} to obtain the K\"ahler forms
\begin{equation}
 J_i={\cal J}_i|_X \eqlabel{Jidef}
\end{equation}
for $i=1,\ldots ,m$ on $X$. In particular, note that these forms are closed, that is
\begin{equation}
  dJ_i=0\quad\mbox{for}\quad i=1,\ldots ,m\; . \eqlabel{Jiprop}
\end{equation}  
There are configurations for which the forms $J_i$ provide a basis for the second cohomology of $X$ and these are sometimes referred to as favourable configurations. In fact, out of the $7890$ configurations provided in the standard list of Ref.~\cite{Candelas:1987kf, Green:1986ck}, some $60\%$ turn out to be favourable in this sense. Furthermore, it has recently been shown~\cite{Anderson:2017aux} that almost all of the other entries in the list have equivalent, favourable configurations. In other words, for almost all of the $7890$ different topological types can a configuration matrix be found such that the forms~\eqref{Jidef} span the entire second cohomology of $X$. This means that the subsequent construction, which will be based on the forms $J_i$, can be thought of as exhausting the entire available space of K\"ahler classes. We recall that the triple intersection numbers $\lambda_{ijk}$ of $X$ can be expressed in terms of the forms $J_i$ by
\begin{equation}
 \lambda_{ijk}=\int_XJ_i\wedge J_j\wedge J_k\; .\eqlabel{isec}
\end{equation} 

As on every Calabi-Yau manifold, we have of course also the holomorphic $(3,0)$ form $\Omega_0$. For CICYs this form can be explicitly constructed~\cite{Candelas:1987kf,Hubsch:1992nu}  by first defining the ambient space $(3,0)$ form $\hat{\Omega}$ via
\begin{equation}
 \hat{\Omega}\wedge dP_1\wedge\cdots\wedge dP_K=\mu\;,\quad \mu=\mu_1\wedge\cdots\wedge\mu_m\;,\quad
 \mu_i=\frac{1}{n_i!}\epsilon_{A_0A_1\cdots A_{n_i}}x_{iA_0}dx_{iA_1}\wedge\cdots\wedge dx_{iA_{n_i}}\; , \eqlabel{Odef0}
\end{equation} 
and then restricting this form to $X$, that is
\begin{equation}
 \Omega_0=\hat{\Omega}|_X\; .\eqlabel{Odef}
\end{equation} 
We note that $\Omega_0$ has the properties
\begin{equation}
 d\Omega_0=0\; ,\qquad \Omega_0\wedge J_i=0\quad\mbox{for}\quad i=1,\ldots ,m\; , \eqlabel{Oprop}
\end{equation} 
the latter as a trivial consequence of the index structure. 

The forms $J_i\wedge J_j\wedge J_k$ as well as $\Omega_0\wedge\bar{\Omega}_0$ are top forms on $X$ and must, therefore, be related by certain functions $\Lambda_{ijk}$ on $X$ such that
\begin{equation}
 J_i\wedge J_j\wedge J_k=\frac{3i}{4}\Lambda_{ijk}\,\Omega_0\wedge\bar{\Omega}_0\; .  \eqlabel{JJJ}
\end{equation} 
We note that integrating this equation over $X$ and using Eq.~\eqref{isec} leads to an alternative expression for the intersection numbers,
\begin{equation}
  \lambda_{ijk}=\int_X\Lambda_{ijk}\Omega_0\wedge\bar{\Omega}_0\; . \eqlabel{isec1}
\end{equation}  
Eq.~\eqref{JJJ} is key in our construction of $SU(3)$ structures on CICY manifolds and generalises the relation~\eqref{relquintic} for the quintic. The computation of the single function $\Lambda_{111}={\cal F}$ in Eq.~\eqref{Fquintic} for the quintic can be generalised, and a general result for $\Lambda_{ijk}$ for an arbitrary configuration matrix can be found.  Since the details are somewhat involved and the general formula turns out to be complicated this calculation has been relegated to Appendix~\secref{appA}. One general rule, which is easily stated, is that $\Lambda_{ijk}=0$ whenever the corresponding triple intersection number $\lambda_{ijk}$ vanishes. For co-dimension one configuration matrices, that is, for $K=1$ and a single defining polynomial $P$, the expression for $\Lambda_{ijk}$ is relatively manageable. In this case, we have for all $i,j,k$ with $\lambda_{ijk}\neq 0$ that
\begin{equation}
 \Lambda_{ijk}= \frac{c_{ijk}}{6\pi^3}\left[\prod_{l=1}^m\frac{|\nabla_l P|^{2n_l}}{\sigma_l}\right](|\nabla_i P|^2|\nabla_j P|^2|\nabla_k P|^2\sigma_i\sigma_j\sigma_k)^{-1}\; , \eqlabel{L1res}
\end{equation}
where $c_{ijk}$ are combinatorial constants and $\nabla_i P$ is the gradient of $P$ with respect to the homogeneous coordinates $x_{iA}$ of the $i^{\rm th}$ projective factor. The combinatorial factors $c_{ijk}$ can be explicitly computed, for example by using Eq.~\eqref{isec1} or from the general expression in Appendix~\secref{appA}. We find that all $c_{ijk}\geq 0$, with equality only if  $\lambda_{ijk}= 0$.

We note that the RHS of Eq.~\eqref{L1res} is homogeneous of degree zero in all coordinates ${\bf x}_i$ and, hence, the $\Lambda_{ijk}$ are indeed well-defined on ${\cal A}$ and on $X$. Further, it follows from Eq.~\eqref{L1res} that the $\Lambda_{ijk}$ are non-singular (since all $\sigma_i$ are non-zero on $\mathbb{P}^{n_i}$) and $\Lambda_{ijk}\geq 0$ everywhere on $X$. These properties of $\Lambda_{ijk}$ in the co-dimension one case are indeed general and also hold for higher co-dimension, as can be seen from the results in Appendix~\secref{appA}.

\subsection[\texorpdfstring{$SU(3)$}{SU(3)} structures on CICY manifolds]{\texorpdfstring{$\boldsymbol{SU(3)}$}{SU(3)} structures on CICY manifolds}
We will now construct $SU(3)$ structures on arbitrary CICY manifolds by specifying a pair $(J,\Omega)$ of a two- and three-form, starting with the Ansatz
\begin{equation}
 J=\sum_{i=1}^m a_i J_i\;,\qquad \Omega=A\,\Omega_0\; , \eqlabel{JOans}
\end{equation} 
where the $(1,1)$-forms $J_i$ and the $(3,0)$-form $\Omega_0$ have been defined in Eqs.~\eqref{Jidef} and \eqref{Odef}, respectively.  Further, the $a_i$ are real, smooth functions on $X$ which are constrained to be strictly positive (so that $J$ is a positive form) but are otherwise arbitrary. The function $A$ on $X$ is real or complex, smooth and should be everywhere non-vanishing.

By virtue of the second relation~\eqref{Oprop} we have $J\wedge\Omega=0$, so that the second requirement for an $SU(3)$ structure in Eq.~\eqref{JOdef} is satisfied independently of the choice of functions $a_i$ and $A$.  Using Eq.~\eqref{JJJ}, we find the first condition~\eqref{JOdef} for an $SU(3)$ structure is satisfied iff
\begin{equation}
 |A|^2=\sum_{i,j,k=1}^m\Lambda_{ijk}a_ia_ja_k\; . \eqlabel{acons}
\end{equation}
Using the explicit expressions for the structure functions $\Lambda_{ijk}$, it can be shown that
\begin{equation}
 \sum_{i,j,k=1}^m\Lambda_{ijk}a_ia_ja_k=|\det(B)|^2\,{\rm det}(g_{\alpha\bar{\beta}})\; , \eqlabel{gres}
\end{equation}
where $g_{\alpha\bar{\beta}}=-2i J_{\alpha\bar{\beta}}$ is the metric associated to $J$ and $\det(B)$, defined in ~\eqref{eq:Omega0Generic}, is the generalization of the factor $|p_{,4}|^2$ that appears in the denominator of $\Omega$ in~\eqref{quinticOmega}.

In conclusion, this means that the forms $(J,\Omega)$ in Eq.~\eqref{JOans} define an $SU(3)$ structure on $X$ iff the functions $a_i$ and $A$ satisfy the constraint~\eqref{acons}. Note that this leaves considerable freedom in the construction. Basically, we can start by choosing any set of real, smooth and strictly positive functions $a_i$, and then use Eq.~\eqref{acons} to define $A$. Since all $\Lambda_{ijk}\geq 0$ and all $a_i\geq 0$, the RHS of Eq.~\eqref{acons} is positive, so $A$ defined in this way can be taken to be real and positive as well. For the quintic, where $m=1$, this leads to a form $J$ given by the conformal re-scaling of a K\"ahler form. In general, for $m>1$, this is no longer necessarily the case, since the functions $a_i$ can be chosen independently. 

Since $dJ_i=d\Omega=0$, it is easy to compute the exterior derivatives of $J$ and $\Omega$ and we find
\begin{equation}
 dJ=\sum_{i=1}^mda_i\wedge J_i\;,\qquad d\Omega=d\ln(A)\wedge\Omega\; .
\end{equation}
Comparison with the general expressions~\eqref{dJdO} for $dJ$ and $d\Omega$ then leads to the torsion classes
\begin{equation}
 W_1=W_2=0\;, \quad W_3=\sum_i(da_i-W_4)\wedge J_i\;,\quad W_4=\frac{1}{2}\sum_i J\hook (da_i\wedge J_i)\;,\quad W_5=d\ln(A)\; .  \eqlabel{torgen}
\end{equation}
Since $W_1$ and $W_2$ vanish, we know from Table~\tabref{ta:manifolds} that there is always a complex structure. The other generic feature of this class of $SU(3)$ structures is that $W_5$ is always an exact one-form. Further details depend on the choice of functions $a_i$. This leaves considerably scope for constructing $SU(3)$ structures based on the Ansatz~\eqref{JOans}, which we only begin to explore in the present paper.

As an example consider the expressions
\begin{equation}
 \sigma_{s,i}:=\sum_{A=0}^{n_i}|x_{iA}|^{2s}\; ,  \eqlabel{si}
\end{equation} 
which are generalisations of $\sigma_i=\sigma_{1,i}$. Since these quantities are nowhere vanishing on ${\cal A}$ and homogeneous of bi-degree $(s,s)$ in $({\bf x}_i,\bar{\bf x}_i)$ they are well-suited to construct smooth, strictly positive functions $a_i$. For example, we can set
\begin{equation}
 a_i=\frac{\sigma_{1,i}^2}{\sigma_{2,i}}\; , \eqlabel{aex}
\end{equation} 
but there are many other choices along similar lines. 

For the remainder of the discussion we will focus on a simple sub-class, characterised by the choice
\begin{equation}
 a_i=a\, t_i\quad\mbox{for}\quad i=1,\ldots ,m\; , \eqlabel{auniv}
\end{equation}
where $a$ is a smooth, strictly positive function on $X$ and the $t_i>0 $ are real constants. In this case, the forms $(J,\Omega)$ can be written as
\begin{equation}
 J=a\,J_0\;,\qquad J_0:=\sum_{i=1}^m t_iJ_i\;,\qquad \Omega=A\,\Omega_0\; . \eqlabel{JOuniv}
\end{equation}
Note that the above $J_0$ is a K\"ahler form and the constants $t_i$ can thus be interpreted as K\"ahler parameters, while $J$ is obtained from $J_0$ by a conformal re-scaling with $a$. Inserting this into Eq.~\eqref{acons}, we find the forms $(J,\Omega)$ in Eq.~\eqref{JOuniv} define an $SU(3)$ structure iff
\begin{equation}
|A|^2=a^3{\cal F}\;,\quad\mbox{where}\quad {\cal F}:=\sum_{i,j,k=1}^m\Lambda_{ijk}t_it_jt_k\; . \eqlabel{aAcons}
\end{equation}
We will refer to this sub-class as ``universal" $SU(3)$ structures. The structure functions $\Lambda_{ijk}$ enter the construction of these universal $SU(3)$ structures only through the function ${\cal F}$, defined above, which can be explicitly computed using the expression~\eqref{L1res} for $\Lambda_{ijk}$ in the case of co-dimension one configurations, or the expression \eqref{eq:LambdaijkHom} in the general case. Specialising Eq.~\eqref{gres} to the universal case we have in particular that
\begin{equation}
 {\cal F}=|\det(B)|^2\, {\rm det}\left(g_{0,\alpha\bar{\beta}}\right)\; ,
\end{equation}
where $g_{0,\alpha\bar{\beta}}=-2iJ_{0,\alpha\bar{\beta}}$ is the metric associated to the K\"ahler form $J_0$ and $B$ is defined in~\eqref{eq:DifferentialsGeneric}. This result shows that ${\cal F}$ is always a strictly positive function (provided all the K\"ahler parameters $t_i>0$). In fact, ${\cal F}$ generalises the function of the same name we have defined for the quintic, see Eq.~\eqref{Fquintic}. 

For universal $SU(3)$ structures, we have the exterior derivatives
\begin{equation}
 dJ=d\ln a\wedge J\;,\qquad d\Omega = d\ln A\wedge \Omega\; ,
\end{equation}
and comparison with Eq.~\eqref{dJdO} shows that the corresponding torsion classes are given by
\begin{equation}
 W_1=W_2=W_3=0\; ,  \quad W_4=d\ln a\;,\quad W_5=d\ln A=\frac{3}{2}d\ln a+\frac{1}{2}d\ln {\cal F}\; .
\end{equation} 
Of course, $W_1$ and $W_2$ are still zero, so that we have a complex structure and $W_5$ remains exact. In addition to the properties of the generic case~\eqref{torgen}, universal $SU(3)$ structures also have  a vanishing $W_3$ torsion class. Also, the class $W_4$  is exact and related to $W_5$ by
\begin{equation}
 W_5=\frac{3}{2}W_4+\frac{1}{2}d\ln {\cal F}\; . \eqlabel{W4W5}
\end{equation} 
The function $a$ is still at our disposal. If it is chosen such that $|d\ln a|\gg |d\ln {\cal F}|$ everywhere on $X$, we have $3W_4\simeq 2W_5$ from Eq.~\eqref{W4W5}. Table~\tabref{ta:manifolds}
shows that this corresponds to an (approximate) conformally Ricci-flat situation, where the conformal factor $a$ dominates over the effect of non Ricci-flatness of the underlying Fubini-Study metric, which causes the appearance of the $d\ln {\cal F}$ term in Eq.~\eqref{W4W5}. In contrast to the exact conformal Calabi-Yau structure normally used in the construction of ${\cal{N}}=1$ type IIB vacua \cite{Giddings:2001yu}, the present approximate structure comes equipped with an explicit, albeit approximate, metric.

Another interesting and obvious choice for $a$ is
\begin{equation}
  a={\cal F}^k\; ,
\end{equation}
for any real number $k$.  This leads to a pattern of torsion classes
\begin{equation}
 W_1=W_2=W_3=0\;,\quad W_4=k\, d\ln {\cal F}\;,\quad W_5=\frac{3k+1}{2}d\ln {\cal F}\; , \eqlabel{W4W51}
\end{equation} 
with $W_4$ and $W_5$ proportional to one another.  For this choice, the limit in which the $SU(3)$ structure becomes approximately conformally Ricci-flat, that is, $3W_4\simeq 2W_5$, can be made more explicit and it corresponds to $k\rightarrow \infty$.

If we set $k=1$ so that 
\begin{equation}
 a={\cal F}\; ,\qquad A={\cal F}^2\; ,\eqlabel{u0}
\end{equation}
then the torsion classes specialise further to
\begin{equation}
 W_1=W_2=W_3=0\;,\quad W_4=d\ln {\cal F}\;,\quad W_5=2\,d\ln {\cal F}\; . \eqlabel{u1}
\end{equation}  
This means that $W_5=2W_4$ and, hence, we have a Strominger-Hull system with $W_3=0$ and a dilaton $\phi$ specified by
\begin{equation}
 d\phi=W_4=d\ln {\cal F}\quad\Rightarrow\quad g_s=e^\phi={\rm const}\times {\cal F}\; . \eqlabel{u2}
\end{equation} 
This shows that, for a suitable choice of integration constant, the string coupling can be kept perturbative. The torsion classes in Eq.~\eqref{u1} represent the direct generalisation of the quintic results to all CICY manifolds. In particular, we find that a Strominger-Hull system can be realised on all CICY manifolds. 

Let us finally remark that the $SU(3)$ structure constructed here does not relate in any obvious way to the unique integrable $SU(3)$ structure that exist on a CY manifold. Naturally, the (3,0)-forms of these two $SU(3)$ structure are necessarily proportional. However, we cannot determine how the Hermitian form $J$ or the K\"ahler forms $J_i$ relates to the K\"ahler form of the integrable $SU(3)$ structure, since the latter is not known explicitly. As mentioned in the Introduction, this ignorance of the integrable K\"ahler form is one of our motivations to construct explicit $SU(3)$ structures on CY manifolds.

\section{Further examples}\seclabel{examples}
While the discussion in the previous section was general and applies to all CICY manifolds it is still useful to work out a few cases other than the quintic more explicitly.  We begin with the two  CICY manifolds which are arguably the simplest after the quintic, the bi-cubic hypersurface in $\mathbb{P}^2\times\mathbb{P}^2$ and the tetra-quadric hypersurface in $\mathbb{P}^1\times\mathbb{P}^1\times\mathbb{P}^1\times\mathbb{P}^1$. Finally, we analyse a more complicated example of a co-dimension two CICY.

\subsection{The bi-cubic}
The bi-cubic hypersurface in the ambient space ${\cal A}=\mathbb{P}^2\times\mathbb{P}^2$  (number 7884 in the list of Ref.~\cite{Candelas:1987kf})  is characterised by the configuration matrix
\begin{equation}
 X\sim\left[\begin{array}{c|c}\mathbb{P}^2&3\\\mathbb{P}^2&3\end{array}\right]^{2,83}_{-162}\qquad
 \begin{array}{l}{\bf x}=(x_0,x_1,x_2)\\{\bf y}=(y_0,y_1,y_2)\end{array}\qquad
 \begin{array}{c}z_1=\frac{x_1}{x_0},z_2=\frac{x_2}{x_0}\\z_3=\frac{y_1}{y_0},z_4=\frac{y_2}{y_0}\end{array}
\end{equation} 
where we have also listed the homogeneous coordinates and the affine coordinates on the patch $U_0$ defined by $x_0\neq 0$, $y_0\neq 0$. The bi-cubic is defined as the zero locus of a bi-cubic polynomial $P=P({\bf x},{\bf y})$, which is related to its affine counterpart $p=p(z_1,\ldots ,z_4)$ on $U_0$ by
\begin{equation}
 p(z_1,\ldots ,z_4)=P(1,z_1,z_2,1,z_3,z_4)\;,\qquad P({\bf x},{\bf y})=x_0^3y_0^3\,p\left(\frac{x_1}{x_0},\frac{x_2}{x_0},\frac{y_1}{y_0},\frac{y_2}{y_0}\right)\; .
\end{equation}
Defining
\begin{equation}
 \begin{array}{r@{~}c@{~}lcr@{~}c@{~}l}
  \sigma_1&=&\sum\limits_{A=0}^2|x_A|^2\;,&& \sigma_2&=&\sum\limits_{A=0}^2|y_A|^2\;,\\[4mm]
 \kappa_1&=&1+|z_1|^2+|z_2|^2\;,&&\kappa_2&=&1+|z_3|^2+|z_4|^2\; ,
\end{array}
\end{equation} 
the two Fubini-Study K\"ahler forms on $U_0$ can be written as
\begin{align}
\begin{split}
 {\cal J}_1&=\frac{i}{2\pi}\partial\bar{\partial}\ln\kappa_1=\frac{i}{2\pi\kappa_1^2}\left[\kappa_1\sum_{a=1}^2dz_a\wedge d\bar{z}_a-\sum_{a,b=1}^2\bar{z}_az_bdz_a\wedge d\bar{z}_b\right]\\
 {\cal J}_2&=\frac{i}{2\pi}\partial\bar{\partial}\ln\kappa_2=\frac{i}{2\pi\kappa_2^2}\left[\kappa_2\sum_{a=3}^4dz_a\wedge d\bar{z}_a-\sum_{a,b=3}^4\bar{z}_az_bdz_a\wedge d\bar{z}_b\right]
 \end{split}
\end{align} 
To restrict these forms to $X$ we can, for example, solve the equation $p=0$ for $z_4=f({\bf z})$, where ${\bf z}=(z_\alpha)=(z_1,z_2,z_3)$ and use
\begin{equation}
 dz_4=-\sum_{\alpha=1}^3\frac{p_{,\alpha}}{p_{,4}}\quad\mbox{where}\quad p_{,i}=\frac{\partial p}{\partial z_i}({\bf z},f({\bf z}))\; .
\end{equation} 
In this way, we can obtain explicit equations for the restricted K\"ahler forms $J_1={\cal J}_1|_X$ and $J_2={\cal J}_2|_X$ as well as for the holomorphic $(3,0)$ form
\begin{equation}
 \Omega_0=\frac{dz_1\wedge dz_2\wedge dz_3}{p_{,4}}\; .
\end{equation}
The only non-vanishing triple intersection numbers of the bi-cubic are $\lambda_{112}=\lambda_{122}=3$ (along with the ones obtained by index permutation) and, hence, we have only two non-vanishing structure functions $\Lambda_{112}$ and $\Lambda_{122}$. They can be computed from the above expressions for $J_i$ and $\Omega_0$. A somewhat tedious calculation shows that
\begin{equation}\begin{array}{r@{~}c@{~}l}
 J_1^2\wedge J_2&=&\frac{3i}{4}\Lambda_{112}\,\Omega_0\wedge\bar{\Omega}_0\;,\qquad \Lambda_{112}=\frac{1}{3\pi^3}\frac{|\nabla_2 P|^2}{\sigma_1^3\sigma_2^2}\\
J_1\wedge J_2^2&=& \frac{3i}{4}\Lambda_{122}\,\Omega_0\wedge\bar{\Omega}_0\;,\qquad \Lambda_{122}=\frac{1}{3\pi^3}\frac{|\nabla_1 P|^2}{\sigma_1^2\sigma_2^3}\; ,
\end{array}\eqlabel{Jibicubic}
\end{equation} 
where $\nabla_1P$ and $\nabla_2 P$ are the gradients of $P$ with respect to the ${\bf x}$ and ${\bf y}$ coordinates, respectively. Note that the above structure functions have the properties mentioned in the general discussion. They are homogeneous of degree zero in both the ${\bf x}$ and ${\bf y}$ coordinates and their complex conjugates and are, hence, well-defined functions on the ambient space $\cal A$ and on the CY $X$. Furthermore, they are non-singular (as $\sigma_1$ and $\sigma_2$ do not vanish) and they are clearly positive everywhere. 

With these ingredients, the construction of $SU(3)$ structures on the bi-cubic proceeds following the general logic explained in the previous section. A pair $(J,\Omega)$ given by the Ansatz
\begin{equation}
 J=\sum_{i=1}^2 a_i J_j\;,\qquad \Omega=A\,\Omega_0
\end{equation}
provides an $SU(3)$ structure iff the constraint
\begin{equation}
 |A|^2=\sum_{i,j,k=1}^2\Lambda_{ijk}a_ia_ja_k\; .
\end{equation}
is satisfied. Inserting the bi-cubic structure functions~\eqref{Jibicubic}, this constraint becomes explicitly
\begin{equation}
 |A|^2=\frac{1}{\pi^3}\left(a_1^2a_2\frac{|\nabla_2 P|^2}{\sigma_1^3\sigma_2^2}+a_1a_2^2\frac{|\nabla_1 P|^2}{\sigma_1^2\sigma_2^3}\right) \eqlabel{consbc}
\end{equation}
Any choice of two smooth and strictly positive functions $a_1$ and $a_2$ on the bi-cubic now leads to an $SU(3)$ structure. We simply insert these two functions into the RHS of Eq.~\eqref{consbc}, which is always strictly positive since $\nabla_1P$ and $\nabla_2P$ cannot vanish simultaneously for a smooth bi-cubic. Then, demanding that $A>0$ everywhere fixes $A$ and, hence, an $SU(3)$ structure whose torsion classes are of the form~\eqref{torgen} and can be explicitly computed from these equations. In this way, choices for $a_i$ such as the ones proposed in Eqs.~\eqref{si} and \eqref{aex}, give rise to a large class of $SU(3)$ structures. 

For the universal case $a_i=a\, t_i$ with $|A|^2=a^3 {\cal F}$, the function ${\cal F}$ reads explicitly
\begin{equation}
{\cal F}= \frac{1}{\pi^3}\left(t_1^2t_2\frac{|\nabla_2 P|^2}{\sigma_1^3\sigma_2^2}+t_1t_2^2\frac{|\nabla_1 P|^2}{\sigma_1^2\sigma_2^3}\right)\; .
\end{equation}
The bi-cubic Strominger-Hull system is then characterised by the general equations~\eqref{u0}--\eqref{u2} with the above functions ${\cal F}$ inserted.

\subsection{The tetra-quadric}\seclabel{TQ}
The tetra-quadric hypersurface in the ambient space ${\cal A}=\mathbb{P}^1\times\mathbb{P}^1\times\mathbb{P}^1\times\mathbb{P}^1$  (number 7862 in the list of Ref.~\cite{Candelas:1987kf}) is described by the configuration
\begin{equation}
 X\sim\left[\begin{array}{c|c}\mathbb{P}^1&2\\\mathbb{P}^1&2\\\mathbb{P}^1&2\\\mathbb{P}^1&2\end{array}\right]^{4,68}_{-128}\qquad
 \begin{array}{l} {\bf x}_1=(x_{1\,0},x_{1\,1})\\{\bf x}_2=(x_{2\,0},x_{2\,1})\\{\bf x}_3=(x_{3\,0},x_{3\,1})\\{\bf x}_4=(x_{4\,0},x_{4\,1})\end{array}\qquad
 \begin{array}{l}z_1=\frac{x_{1\,1}}{x_{1\,0}}\\z_2=\frac{x_{2\,1}}{x_{2\,0}}\\z_3=\frac{x_{3\,1}}{x_{3\,0}}\\z_4=\frac{x_{4\,1}}{x_{4\,0}}\end{array}
\end{equation} 
where the homogeneous coordinates for each $\mathbb{P}^1$ factor and their affine counterparts on the patch $U_0=\{x_{i,0}\neq 0\}$ are listed on the right. The tetra-quadric is defined as the zero locus of a polynomial $P=P({\bf x}_1,\ldots ,{\bf x}_4)$ of multi-degree $(2,2,2,2)$, related to its affine counterpart $p=p(z_1,\ldots ,z_4)$ by
\begin{equation}
 p({\bf z})=P(1,z_1,1,z_2,1,z_3,1,z_4)\;,\quad P({\bf x}_1,{\bf x}_2,{\bf x}_3,{\bf x}_4)=\left(\prod_{i=1}^4x_{i\,0}^2\right)p\left(\frac{x_{1\,1}}{x_{1\,0}},\frac{x_{2\,1}}{x_{2\,0}},\frac{x_{3\,1}}{x_{3\,0}},\frac{x_{4\,1}}{x_{4\,0}}\right)\; .
\end{equation} 
As usual we define 
\begin{equation}
 \sigma_i=\sum_{A=0}^1|x_{iA}|^2\;,\qquad \kappa_i=1+|z_i|^2\quad\mbox{for}\quad i=1,\ldots ,4\;,
\end{equation} 
which leads to the Fubini-Study K\"ahler forms
\begin{equation}
 {\cal J}_i=\frac{i}{2\pi}\partial\bar{\partial}\ln\kappa_i=\frac{i}{2\pi}\frac{dz_i\wedge d\bar{z}_i}{\kappa_i^2}\quad\mbox{for}\quad i=1,\ldots, 4\; .
\end{equation} 
We can restrict these forms to the tetra-quadric, $J_i={\cal J}_i|_X$, by solving, for example, for $z_4=f({\bf z})$, where ${\bf z}=(z_\alpha)=(z_1,z_2,z_3)$ and use
\begin{equation}
 dz_4=-\sum_{\alpha=1}^3\frac{p_{,\alpha}}{p_{,4}}\quad\mbox{where}\quad p_{,i}=\frac{\partial p}{\partial z_i}({\bf z},f({\bf z}))\; .
\end{equation} 
On the patch $U_0$, this leads to
\begin{equation}
 J_\alpha=\frac{i\,dz_\alpha\wedge d\bar{z}_\alpha}{2\pi \kappa_\alpha^2}\;,\qquad
 J_4=\frac{i}{2\pi \kappa_4^2}\sum_{\alpha,\beta=1}^3v_\alpha\bar{v}_\beta\, dz_\alpha\wedge d\bar{z}_\beta\quad\mbox{with}\quad v_\alpha:=\frac{p_{,\alpha}}{p_{,4}}\; , \eqlabel{JTQ}
\end{equation} 
while the holomorphic $(3,0)$ form on $X$ can be written as
\begin{equation}
 \Omega_0=\frac{dz_1\wedge dz_2\wedge dz_3}{p_{,4}}\; .
\end{equation}
The only non-vanishing triple intersection numbers of the tetra-quadric are $\lambda_{123}=\lambda_{124}=\lambda_{134}=\lambda_{234}=2$ and, by inserting the above expressions for $J_i$ and $\Omega_0$ into Eq.~\eqref{JJJ}, we find for the corresponding structure functions
\begin{equation}
  \Lambda_{ijk}=\frac{1}{6\pi^3}\frac{|\nabla_lP|^2\sigma_l}{\sigma_l^2\sigma_2^2\sigma_3^2\sigma_4^2}=\frac{1}{6\pi^3}\frac{|p_l|^2\kappa_l^2}{\kappa_1^2\kappa_2^2\kappa_3^2\kappa_4^2}=:\frac{1}{6}\Lambda_l \quad\mbox{where}\quad \{i,j,k,l\}=\{1,2,3,4\}\; , \eqlabel{LTQ}
\end{equation}
and permutations thereof. All other $\Lambda_{ijk}$ vanish. We note that, in line with our general statements, all $\Lambda_{ijk}$ are well-defined, smooth and positive. 

The Ansatz
\begin{equation}
 J=\sum_{i=1}^4a_iJ_i\;,\qquad \Omega=A\,\Omega_0\; ,
\end{equation}
then leads to an $SU(3)$ structure $(J,\Omega)$ on the tetra-quadric iff $|A|^2=\sum_{i,j,k=1}^4\Lambda_{ijk}a_ia_ja_k$ and, with the above structure functions~\eqref{LTQ}, this condition turns into
\begin{equation}
 |A|^2=a_1a_2a_3a_4\sum_{i=1}^4a_i^{-1}\Lambda_i=\frac{a_1a_2a_3a_4}{\pi^3\kappa_1^2\kappa_2^2\kappa_3^2\kappa_4^2}\sum_{i=1}^4a_i^{-1}\kappa_i^2|p_{,i}|^2\; .
\end{equation}
From this equation, any choice of four smooth functions $a_i>0$ on $X$ determines a function $A>0$ and, hence, an $SU(3)$ structure. As is clear from the examples described by Eqs.~\eqref{si} and \eqref{aex}, there is considerable freedom in this construction.

For the universal case $a_i=a\,t_i$, the function $A$ is determined by $|A|^2=a^3{\cal F}$, where ${\cal F}$ is explicitly given by
\begin{equation}
 {\cal F}=t_1t_2t_3t_4\sum_{i=1}^4t_i^{-1}\Lambda_i=\frac{t_1t_2t_3t_4}{\pi^3\kappa_1^2\kappa_2^2\kappa_3^2\kappa_4^2}\sum_{i=1}^4t_i^{-1}\kappa_i^2|p_{,i}|^2\; . \eqlabel{FTQ}
\end{equation} 
For a smooth tetra-quadric not all $\Lambda_i$ vanish simultaneously, so ${\cal F}>0$ everywhere on $X$. 

The Strominger-Hull system on the tetra-quadric is described by the general equations~\eqref{u0}--\eqref{u2} with the above function ${\cal F}$ inserted. 

\subsection{A co-dimension two CICY}
\seclabel{sec:CICY7888}
The purpose of this example is two-fold. First, we show how to generalise the methods from the co-dimension one case to higher co-dimensions. Second, we illustrate how this leads to the general formula for the structure functions $\Lambda_{ijk}$ given in Appendix~\secref{appA}. The example we are working with is a co-dimension two CICY (number 7888 in the list of Ref.~\cite{Candelas:1987kf}) in the projective ambient space $\mathbb{P}^1\times\mathbb{P}^4$, with configuration matrix
 \begin{equation}
 X\sim\left[\begin{array}{c|cc}\mathbb{P}^1&0&2\\\mathbb{P}^4&4&1\end{array}\right]^{2,86}_{-168}~~~
 \begin{array}{l}{\bf x_1}=(x_{1\,0},x_{1\,1})\\{\bf x_2}=(x_{2\,0},\ldots, x_{2\,4})\end{array}~~~
 \begin{array}{l}z_{1}=\frac{x_{1\,1}}{x_{1\,0}}\\z_{2}=\frac{x_{2\,1}}{x_{2\,0}},\ldots,z_5=\frac{x_{2\,5}}{x_{2\,0}}\end{array}\; .
\end{equation} 
We denote the two defining equations by $P_u$ and their affine counterparts in the patch $U_0=\{x_{10}\neq 0,\,x_{20}\neq 0\}$ by $p_u$, where $u=1,2$. The explicit computation proceeds similarly to the co-dimension one case. First, we use the equations $p_u=0$ to solve for, say, $z_4$ and $z_5$ in terms of ${\bf z}=(z_1,z_2,z_3)=(z_\alpha)$, and replace their differentials $dz_{4}$ and $dz_5$ using
\begin{align}
\begin{pmatrix}
p_{1,4} & p_{1,5}\\[3mm]
p_{2,4} &p_{2,5} 
\end{pmatrix}
\begin{pmatrix}
d z_4\\[4mm]
d z_5
\end{pmatrix}
=
-\begin{pmatrix}
\sum_{\alpha=1}^3 p_{1,\alpha} d z_\alpha\\[3mm]
\sum_{\alpha=1}^3 p_{2,\alpha} d z_\alpha
\end{pmatrix}\;, \eqlabel{eq:BijkExplicitCD2}
\end{align}
where $p_{u,i}=\partial P_u/\partial z_i$. Denoting the matrix in this equation by $B$, we use
\begin{align}
B^{-1}=
\frac{1}{\det(B)}
\begin{pmatrix}
\phantom{-}p_{2,5}  & -p_{1,5}\\[3mm]
-p_{2,4} & \phantom{-}p_{1,4}
\end{pmatrix}
\end{align}
to arrive at
\begin{align}
\begin{split}
d z_4 &= -\frac{1}{\det(B)}\sum_{i=1}^3\left(p_{2,5}\; p_{1,i} - p_{1,5}\; p_{2,i}  \right)dz_i\\
d z_5 &= -\frac{1}{\det(B)}\sum_{i=1}^3\left(p_{1,4}\; p_{2,i} - p_{2,4}\; p_{1,ii}  \right)dz_i\;.
\end{split}
\eqlabel{eq:ExampleDifferentials}
\end{align}
In affine coordinates, the two Fubini-Study K\"ahler forms are given by
\begin{align}
\begin{split}
 {\cal J}_1&=\frac{i}{2\pi}\partial\bar{\partial}\ln\kappa_1=\frac{i}{2\pi\kappa_1^2}[dz_1\wedge d\bar{z}_1]\;,\\
 {\cal J}_2&=\frac{i}{2\pi}\partial\bar{\partial}\ln\kappa_2=\frac{i}{2\pi\kappa_2^2}\left[\kappa_2\sum_{\alpha=2}^3dz_\alpha\wedge d\bar{z}_\alpha-\sum_{\alpha,\beta=2}^3\bar{z}_\alpha z_\beta dz_\alpha\wedge d\bar{z}_\beta\right]\;,
 \end{split}
\end{align} 
where
\begin{equation}
 \kappa_1=1+|z_1|^2\;,\qquad \kappa_2=1+\sum_{a=2}^5|z_a|^2\; .
\end{equation} 
As a next step we need to compute the holomorphic three-form $\Omega_0$. Applying its general definition~\eqref{Odef0}, \eqref{Odef} to the present case leads to
\begin{align}
\Omega_0=\frac{dz_1\wedge dz_2\wedge dz_3}{p_{1,4}\;p_{2,5}-p_{1,5}\;p_{2,4}}\;.
\end{align}
The only non-vanishing intersection numbers of this CICY are $\lambda_{122}=4$ and $\lambda_{222}=8$ and the corresponding non-vanishing structure functions can be written in the form
\begin{align}
\eqlabel{eq:CICY7888Lambda122}
\begin{split}
\Lambda_{122}&=\frac{1}{6\pi^3 \kappa_1^2\kappa_2^3}\sum _{a,b,c,d=2}^5 \left(\delta _{a,b} \delta _{c,d}+ z_a \bar{z}_b \delta _{c,d}+z_c \bar{z}_d \delta _{a,b}+z_a \bar{z}_b z_c \bar{z}_d\right)\times\\ 
&\hspace{3.5cm}\left(p_{1,c} \bar{p}_{1,d} p_{2,a} \bar{p}_{2,b} - p_{1,a} \bar{p}_{1,d} p_{2,c}\bar{p}_{2,b} - p_{1,c}\bar{p}_{1,b}p_{2,a} \bar{p}_{2,d}+p_{1,a} \bar{p}_{1,b} p_{2,c} \bar{p}_{2,d}\right)\\
\Lambda_{222}&=\frac{1}{\pi^3 \kappa_2^4}\sum _{a,b=2}^5~\sum _{c,d=1}^1\left(\delta _{a,b} \delta _{c,d}+ z_a \bar{z}_b \delta _{c,d}\right)\times\\ 
&\hspace{3.5cm}\left(p_{1,c} \bar{p}_{1,d}p_{2,a} \bar{p}_{2,b}  - p_{1,a}\bar{p}_{1,d} p_{2,c} \bar{p}_{2,b} - p_{1,c} \bar{p}_{1,b} p_{2,a} \bar{p}_{2,d}+p_{1,a} \bar{p}_{1,b} p_{2,c} \bar{p}_{2,d}\right)
\end{split}
\end{align}
Finally, we compare this with the result from the general expression \eqref{eq:AffineLambdaijkGeneral}: 
\begin{enumerate}
\item The matrix $B$ in \eqref{eq:BijkExplicitCD2} is nothing but $B_2[1,2,3]$ and \eqref{eq:ExampleDifferentials} are the differentials obtained from~\eqref{eq:DifferentialsGeneric}.
\item For $\Lambda_{122}$, the sums are all symmetric and hence every term occurs twice. That means that upon factoring out the common factor of 2 from the numerator, we get a factor of $1/3$ in the denominator, matching the symmetry factors $c_{ijk}$ of~\eqref{eq:cijkSymmetryFactors} that appear in~\eqref{eq:AffineLambdaijkGeneral}.
\item We always have to skip three columns in the Jacobian. In $\Lambda_{122}$ the index structure shows that we skip the column which contains derivatives w.~r.~t.~the $\mathbb{P}^1$ coordinate $z_1$, which is why the sums run from 2 to 5. Likewise, in $\Lambda_{222}$ we delete columns such that a $\mathbb{P}^1$ coordinate and a $\mathbb{P}^4$ coordinate are left, so one sum runs over the $\mathbb{P}^1$ block, i.e.\ from 1 to 1, and the other over the $\mathbb{P}^4$ block, that is, again from 2 to 5.
\item The indices on $p_{u,s}$ show anti-symmetry structures which arise from a determinant of the Jacobian with three columns deleted.
\item The terms $z_a \bar{z}_b z_c \bar{z}_d$ actually sum to zero due to the symmetry/antisymmetry structure of the indices.
\item The expression has the structure one would expect from the terms $(\tilde\nabla_{i_1}p_{u_1}\cdot\tilde\nabla_{\bar{\imath}_1}\bar{p}_{\bar{u}_{1}})(\tilde\nabla_{i_2}p_{u_2}\cdot\tilde\nabla_{\bar{\imath}_2}\bar{p}_{\bar{u}_{2}})$ in \eqref{eq:AffineLambdaijkGeneral}, where we have carried out the anti-symmetrisation from the $\epsilon_{u_{1}u_{2}}$ factors.
\end{enumerate}
For the universal case $a_i=a\,t_i$, the function $A$ is determined by $|A|^2=a^3{\cal F}$, where ${\cal F}$ is obtained by inserting the structure functions~\eqref{eq:CICY7888Lambda122} into the second Eq.~\eqref{aAcons}. For the Strominger-Hull system on this co-dimension two CICY, the function ${\cal F}$ is used in the general equations~\eqref{u0}--\eqref{u2}. 

\section{The Bianchi identity}\seclabel{BI}
We have seen that large classes of non-trivial $SU(3)$ structures can be constructed on CICY manifolds. In particular, we have demonstrated the existence of a Strominger-Hull system on every CICY manifold. These $SU(3)$ structures are potentially relevant for string theory, both in the context of heterotic and type II compactifications. While we leave a comprehensive study of the relevant applications in string theory for future work, we present here a brief initial discussion of the Bianchi identities for the anti-symmetric tensor fields. These identities need to be satisfied in order to obtain a full string solution and this proves a somewhat difficult task. Consequently, we limit our discussion to some general remarks and then re-visit the tetra-quadric example.

\subsection{Generalities}
We recall that a Strominger-Hull $SU(3)$ structure $(J,\Omega)$ on a CICY manifold $X$ was obtained by setting
\begin{equation} 
 J={\cal F} J_0\;,\quad \Omega= {\cal F}^2\Omega_0\;, \quad J_0=\sum_{i=1}^mt_iJ_i\; , \eqlabel{JO1}
\end{equation}
where $J_i={\cal J}_i|_X$ are the ambient space K\"ahler forms restricted to $X$ and $\Omega_0$ is the holomorphic $(3,0)$ form on $X$. The constants $t_i>0$ can be interpreted as the analogue of K\"ahler parameters and the function ${\cal F}$ can be written as
\begin{equation}
 {\cal F}=\sum_{i,j,k=1}^m\Lambda_{ijk}t_it_jt_k\; ,
\end{equation} 
where the structure functions have been computed for all CICY manifolds in Eq.~\eqref{L1res} and Appendix~\secref{appA}. It turns out that for a smooth manifolds, ${\cal F}>0$ everywhere on $X$ as must be the case in order for $J$ to be a positive form.

Since $dJ_i=0$ and $d\Omega_0=0$, the exterior derivatives of $J$ and $\Omega$ are easily computed, and lead to the torsion classes
\begin{equation}
 W_1=W_2=W_3=0\;,\qquad W_5=2W_4=2\,d\ln {\cal F}\; . \eqlabel{sht}
\end{equation}  
The dilaton $\phi$, both in the context of type II and heterotic string theory, satisfies
\begin{equation}
 d\phi=W_4=d\ln {\cal F}\;, \eqlabel{dil1}
\end{equation} 
so that the string coupling $g_s=e^\phi={\rm const}\times {\cal F}$ can be kept perturbative, for a suitable choice of the integration constant. 

The final ingredient required for this to become a string solution is that the torsion~\eqref{sht} is supported by a suitably matching flux. Let us focus on the case where this flux is the NS three-form field strength $H$, either in the heterotic or type II context. In either case,  $H$ is given by Eq.~\eqref{HSH} and, inserting $J$ from Eq.~\eqref{JO1} leads to
\begin{equation}
 H=i(\partial-\bar{\partial})J=i\left(\partial \ln {\cal F}-\bar{\partial}\ln {\cal F}\right)\wedge J=i(\partial {\cal F}-\bar{\partial} {\cal F})\wedge J_0\; .
\end{equation} 
This is the NS flux we need to add to the compactification in order to obtain a string solution. It is important to note that $H$ is not closed, 
\begin{equation}
 dH= 2i\frac{\bar{\partial}\partial {\cal F}}{{\cal F}}\wedge J=2i\,\bar{\partial}\partial {\cal F}\wedge J_0 \; , \eqlabel{dHres}
\end{equation} 
and, hence, this flux needs to be supported by a non-zero source term in the Bianchi identity. In type II string theory this requires an NS5 brane source equal to the RHS of Eq.~\eqref{dHres} (or a D5 brane source in the case where $H$ is interpreted as a RR flux). We recall that by equation \eqref{JO1}, ${\cal F}$ must be a smooth function, and so the RHS of  \eqref{dHres} cannot be proportional to a $\delta$ function. In other words, the source required to satisfy the equation requires some smearing.\footnote{It is possible that non-universal $SU(3)$ structures, with $J = \sum_i a_i J_i$, give more freedom to engineer solutions with localized sources.}  We leave the subtleties of such constructions for future work, but remark that sources of this type have been discussed for vacua on solvmanifolds in Ref.~\cite{Andriot:2015sia}. Related discussions can also be found for example in Refs.~\cite{Gauntlett:2003cy,Kachru:2002sk,Schulz:2004ub,Caviezel:2008ik,McOrist:2012yc,Petrini:2013ika}. It would be interesting to explore if the methods used in these references  can be useful to find the required source term. 

In the heterotic case, the Bianchi identity reads (to leading order in the $\alpha'$ expansion)\footnote{{As is the case for type II vacua, the Bianchi identity can be modified by source terms related to (smeared) NS5 branes. We leave the discussion of this possibility for future work.}}
\begin{equation}
d H=\tfrac{\alpha'}{4}\left({\rm tr}(F\wedge F)-{\rm tr}( R\wedge R)\right)\:,
\end{equation}
and the task is to match this to the RHS of Eq.~\eqref{dHres}. Since we have specified a metric, the ${\rm tr}(R\wedge R)$ term can be computed explicitly. Hence, solving the Bianchi identities reduces to the task of finding a suitable vector bundle $V\rightarrow X$ with a connection $A$ and associated field strength $F$ such that 
\begin{equation}
2i\,\bar{\partial}\partial {\cal F}\wedge J_0=\tfrac{\alpha'}{4}\left({\rm tr}(F\wedge F)-{\rm tr}( R\wedge R)\right)\:. \eqlabel{BI2}
\end{equation}
Note that the gauge connection also has to satisfy the supersymmetry conditions~\eqref{Fsusy} at the same time. This appears to be a difficult task as there is no obvious guidance for the choice of a suitable bundle $V$ and its connection. In the present paper, we will not attempt to construct such bundles and connections.

Instead, we compute the term ${\rm tr}(R\wedge R)$ in order to gain a clearer picture which contribution from ${\rm tr}(F\wedge F)$ is required. Of considerable help for this task is the observation is that ${\rm tr}(R\wedge R)$ is invariant under a conformal re-scaling of the metric~\cite{Avez:70}. This means that, instead of using the metric $g$ associated to the $SU(3)$ structure~\eqref{JO1}, we can use the K\"ahler metric $g_0$ associated to the K\"ahler form $J_0$. Despite this simplification, it seems difficult to express ${\rm tr}(R\wedge R)$ in a manageable and suggestive form in complete generality for any CICY manifold. For this reason, we will compute ${\rm tr}(R\wedge R)$ for an example, namely the tetra-quadric CICY discussed in Section~\secref{TQ}.

\subsection{The tetra-quadric re-visited}
Our task is to compute the ${\rm tr}(R\wedge R)$ term in the Bianchi identity for the case of the tetra-quadric with the metric associated to the Strominger-Hull $SU(3)$ structure~\eqref{JO1}.
As discussed, this can be done in terms of the K\"ahler metric $g_0$, associated to the K\"ahler form $J_0=\sum_{i=1}^4t_iJ_i$. Summing up the explicit forms $J_i$ for the tetra-quadric in Eq.~\eqref{JTQ} gives
\begin{equation}
 J_0=\partial\bar{\partial}K=\frac{i}{2\pi}\sum_{\alpha,\beta=1}^3\left[t_\alpha\kappa_\alpha^{-2}\delta_{\alpha\beta}+t_4\kappa_4^{-2}v_\alpha\bar{v}_\beta\right]dz^\alpha\wedge d\bar{z}^\beta\; ,
\end{equation}
where $v_\alpha=p_{,\alpha}/p_{,4}$ and the K\"ahler potential is given by $K=\frac{i}{2\pi}\sum_{i=1}^4t_i\ln\kappa_i$, restricted to $X$. 
The associated metric $g_{0,\alpha\bar{\beta}}=-2iJ_{0,\alpha\bar{\beta}}$ and its inverse $g^{\alpha\bar{\beta}}_0$ can then be written as
\begin{equation}
g_{0,\alpha\bar{\beta}}=\frac{1}{\pi}\left[t_\alpha\kappa_\alpha^{-2}\delta_{\alpha\bar{\beta}}+t_4\kappa_4^{-2}v_\alpha\bar{v}_{\bar{\beta}}\right]\;,\qquad
g^{\alpha\bar{\beta}}_0=\pi\left[t_\alpha^{-1}\kappa_\alpha^2\delta^{\alpha\bar{\beta}}-t_4^{-1}\kappa_4^2w^\alpha\bar{w}^{\bar{\beta}}\right]\;,
\end{equation}
where we have introduced the short-hand notation
\begin{equation}
 w^\alpha=\lambda r_\alpha^2\bar{v}^\alpha\;,\qquad r_\alpha^2=\frac{t_4\kappa_\alpha^2}{t_\alpha\kappa_4^2}\;,\qquad \lambda^{-2}=1+\sum_{\alpha=1}^3r_\alpha^2|v_\alpha|^2\; .
\end{equation} 
A useful relation between those quantities is
\begin{equation}
 \sum_{\alpha=1}^3v_\alpha w^\alpha=\lambda^{-1}-\lambda\; .
\end{equation} 
In order to compute the curvature two-form, we use the standard K\"ahler geometry relations
\begin{equation}
 R_\alpha^\beta=-\bar{\partial}\Gamma_\alpha^\beta\;,\qquad \Gamma_\alpha^\beta=g^{\beta\bar{\gamma}}_0\partial g_{0,\alpha\bar{\gamma}}\; ,
\end{equation} 
and inserting the above metric and its inverse leads to the connection one-form
\begin{equation}
 \Gamma_\alpha^\beta=-2\partial\ln\kappa_\alpha\delta_\alpha^\beta+\frac{\lambda^2r_\beta^2}{r_\alpha^2}\partial(r_\alpha^2v_\alpha\bar{v}^\beta)\; , 
\end{equation} 
and the curvature two-form
\begin{equation}
 R_\alpha^\beta=-4\pi iJ_\alpha\delta_\alpha^\beta-\Omega_\alpha^\beta\;,\qquad \Omega_\alpha^\beta=\bar{\partial}\left(\frac{\lambda^2r_\beta^2}{r_\alpha^2}\partial(r_\alpha^2v_\alpha\bar{v}^\beta)\right)\;.\label{Rres}
\end{equation}
As a useful crosscheck of our calculation, it is straightforward to check that this expression leads to a Ricci-form ${\cal R}=\sum_{\alpha=1}^3 R_\alpha^\alpha$ which satisfies
\begin{equation}
{\cal  R}= \partial \bar{\partial}  \ln \det g_0\; ,
\end{equation}
as required for a K\"ahler metric.

The desired quantity ${\rm tr}(R\wedge R)$ can then be written as
\begin{equation}
 {\rm tr}(R\wedge R)=\sum_{\alpha, \beta=1}^3 (R_\alpha^\beta \wedge R_\beta^\alpha)+\mbox{c.c.}=8\pi i\sum_{\alpha=1}^3 J_\alpha\wedge\Omega_\alpha^\alpha+\sum_{\alpha,\beta=1}^3\Omega_\alpha^\beta\wedge\Omega_\beta^\alpha+\mbox{c.c.}\; . \eqlabel{RRTQ}
\end{equation} 
It is interesting that this result can be expressed in terms of the tetra-quadric structure functions $\Lambda_i$ given in Eq.~\eqref{LTQ} and the function ${\cal F}$ in Eq.~\eqref{FTQ} by writing
\begin{equation}
 \Omega_\alpha^\beta=\frac{v_\alpha}{v_\beta}\tilde{\Omega}_\alpha^\beta\; ,\qquad
  \tilde{\Omega}_\alpha^\beta=\frac{t_1t_2t_3t_4}{t_\beta}\bar{\partial}\left(\frac{\Lambda_4\Lambda_\beta}{{\cal F}\Lambda_\alpha}\partial\left(\frac{\Lambda_\alpha}{\Lambda_4}\right)\right)\; .
\end{equation}
Note that when inserted into Eq.~\eqref{RRTQ}, the pre-factor $v_\alpha/v_\beta$ drops out so that ${\rm tr}(R\wedge R)$ only depends on $\tilde{\Omega}_\alpha^\beta$ and the K\"ahler forms $J_\alpha$. It is encouraging that ${\rm tr}(R\wedge R)$ depends on the same quantities as $dH$. We hope this result will ultimately provide some guidance as to which vector bundle and connection to choose for the tetra-quadric, in order to satisfy the Bianchi identity.

\section{Conclusion}\seclabel{conclusion}
Compactification on Calabi-Yau three-folds and on manifolds with $SU(3)$ structure are usually seen as complementary within string theory. In this paper, we have explored the relation between these two approaches by constructing non-trivial $SU(3)$ structures on Calabi-Yau three-folds and by analysing their possible role in string compactifications.

Focusing on the relatively easily accessible complete intersection Calabi-Yau manifolds in product of projective spaces (CICY manifolds) we have obtained a number of interesting results. We have seen that, using the K\"ahler forms provided by the projective ambient spaces and the holomorphic $(3,0)$ form available on a CY manifold as basic building blocks, we can obtain large classes of $SU(3)$ structures on all CICY manifolds. For our construction, all these $SU(3)$ structures have vanishing torsion classes $W_1$ and $W_2$ and, hence, have an associated complex structure. The other three torsion classes $W_3$, $W_4$ and $W_5$ are generically non-zero and are given in terms of a set of smooth, strictly positive but otherwise arbitrary functions $a_i$ on the CICY manifold. In a ``universal" case, where all functions $a_i$ are proportional, we obtain a Strominger-Hull system with $W_3=0$ on every CICY manifold. 

Such $SU(3)$ structures can lead to string backgrounds in both heterotic and type II string theory with a non-trivial dilaton profile and non-vanishing NS flux. We have computed the dilaton profile required for the Strominger-Hull system on CICY manifolds and find that the string coupling can be kept in the perturbative range. Further, we have determined the NS flux  $H$ supporting this solution and it turns out that it is not closed and, hence, needs to be supported by sources in the Bianchi identity. In the type II case, this requires (smeared) NS five-brane solutions and in the heterotic case a suitable vector bundle. 

We have explicitly discussed a number of examples, namely the quintic in $\mathbb{P}^4$, the bi-cubic hypersurface in $\mathbb{P}^2\times\mathbb{P}^2$, the tetra-quadric hypersurface in $\mathbb{P}^1\times\mathbb{P}^1\times\mathbb{P}^1\times\mathbb{P}^1$ and a co-dimension two CICY manifold. For the tetra-quadric, we have taken preliminary steps towards solving the Bianchi identity by computing the ${\rm tr}(R\wedge R)$ term. 

A number of follow-up questions and further directions are suggested by the results in this paper. First and foremost, it would be desirable to clarify, in the case of the universal Strominger-Hull system, whether the Bianchi identity in the type II or heterotic case can be solved. Only if this can be accomplished have we found a proper string solution. Our construction also leads to a large class of non-universal $SU(3)$ structures, where the functions $a_i$ are not proportional to one another, and these should be explored in more detail. It is possible that other $SU(3)$ structures of interest to string theory can be found among those non-universal cases. 

Moreover, while we have primarily manipulated the real two-form $J$ of the $SU(3)$ structure in this paper, there is scope to construct interesting non-trivial $SU(3)$ structures by a less restricted Ansatz for the complex three-form $\Omega$. Such generalisations are necessary in order to produce $SU(3)$ structures with non-integrable almost complex structure. It would, for example, be interesting to explore if type IIA $SU(3)$ vacua can be constructed in this way.\footnote{{Recently, it has been shown in Ref.~\cite{Terrisse:2017tbb}, that such type IIA $SU(3)$ vacua are allowed on manifolds that are $\mathbb{CP}^1$ fibrations over K\"ahler-Einstein 4-manifolds.}}

While our construction was carried out for CICY manifolds, it likely generalises to other classes of CY three-folds, notably to CY three-folds constructed as hypersurfaces in toric four-fold ambient spaces~\cite{Kreuzer:2000xy}. There is indeed overlap between CICY three-folds and CY hypersurfaces in toric four-folds and some of the examples considered in this paper, specifically the quintic, the bi-cubic and the tetra-quadric appear in both lists. 

Finally, our construction should readily generalise to CY four-folds, in this case leading to $SU(4)$ structures. For example, a suitable modification of the quintic example could be applied to the sextic in $\mathbb{P}^5$ or adding another $\mathbb{P}^1$ factor to the tetra-quadric example would lead to $SU(4)$ structures on the degree $(2,2,2,2,2)$ CY hypersurface in $(\mathbb{P}^1)^5$.  It would be interesting to study this generalisation and its possible applications to F-theory compactifications.

\section*{Acknowledgments}
We thank {David Andriot, Callum Brodie, Ulf Danielsson, James Gray, George Papadopoulos, Achilleas Passias, Dimitrios Tsimpis} and Daniel Waldram for useful discussions. We would also like to thank James Gray for collaboration on an earlier, related project.
M.~L.~is financed by the Swedish Research Council (VR) under grant number 2016-03873. A.~L.~and F.~R.~are supported by the EPSRC network grant EP/N007158/1. The work of A.~L.~is furthermore partially supported by the EC 6th Framework Programme MRTN-CT-2004-503369. 

\appendix

\section{Some technical results} \seclabel{appA}
In this appendix we present a general formula for the structure function of any CICY configuration matrix, including the 7890 specific configuration matrices from Ref.~\cite{Candelas:1987kf}. 

Recall that the ambient space is a product ${\cal A}=\mathbb{P}^{n_1}\times\cdots\times\mathbb{P}^{n_m}$ of $m$ projective spaces, each with dimension $n_i$ and with total dimension $d=\sum_{i=1}^mn_i$. We start by computing the affine version of the structure functions in the patch $U_0=\{x_{i0}\neq 0\}$, before giving the homogeneous version. The coordinates on this patch are given by ${\bf z}_i=(z_{i1},\ldots ,z_{in_i})$, where $i=1,\ldots ,m$. We denote the affine versions of the defining polynomials by $p_u$, where $u=1,\ldots ,K$ and $K=d-3$. In order to avoid double indices, we also denote the affine coordinates collectively by $z_s$, where $s=1,\ldots ,d$. We introduce the $K\times d$ Jacobi matrix
\begin{align}
A_u^{s}=\left(\frac{\partial p_{u}}{\partial z_s}\right)=:p_{u,s}\;,
\end{align}
where we define $p_{u,s}=\partial p_{u}/\partial z_s$. By deleting three columns $r,s,t$ from $A$ we obtain $K\times K$ matrices denoted by $B_K[r,s,t]$ and, for ease of notation, we also introduce associated index sets  $\mathcal{I}[r,s,t]=\{1,\ldots ,d\}\backslash\{r,s,t\}$. With this notation, we can write
\begin{align}
d z_v =-\sum_u \sum_{\ell\in\{r,s,t\}}  (B^{-1}_K[r,s,t])^v_{u}\; p_{u,\ell}\; dz_\ell\;.
\eqlabel{eq:DifferentialsGeneric}
\end{align}
Using $B_K[r,s,t]$, we can generalise the expression~\eqref{quinticOmega} for the holomorphic $(3,0)$-form $\Omega_0$ to
\begin{align}
\eqlabel{eq:Omega0Generic}
\Omega_0=\frac{dz_r\wedge dz_s\wedge dz_t}{\det(B_K[r,s,t])}\;.
\end{align}

Lastly, in order to describe the general expression, we define for each index $s=(ia)$ the $(n_i+1)$-dimensional gradients
\begin{align}
\tilde\nabla_{s} p_u=\tilde\nabla_{ia} p_u=\left(p_{u,i1},~p_{u,i2},~\ldots~,~p_{u,in_i},-\sum_{a=1}^{n_i}z_{ia}\; p_{u,ia}\right)\;,
\end{align}
where $p_{u,ia}=\partial p_u/\partial z_{ia}$. It should be noted that the $\tilde\nabla_{ia}$ do in fact not depend explicitly on the index $a$, and hence they are the same for all coordinates $z_{ia}$ from the same ambient space factor $i$, that is, $\tilde\nabla_{i1}p_u=\ldots=\tilde\nabla_{in_i}p_u$. It turns out that this somewhat redundant definition is helpful in order to incorporate certain combinatorial factors in the final expression for the $\Lambda_{ijk}$. Given this notation, a straightforward but tedious computation based on Eq.~\eqref{JJJ} leads to
\begin{align}
\eqlabel{eq:AffineLambdaijkGeneral}
\begin{split}
\Lambda_{ijk}=&\frac{c_{ijk}}{\pi^3}\left(\frac{1}{\kappa_i\kappa_j\kappa_k}\prod_{l=1}^{m}\frac{1}{\kappa_l}\right)\times\\&\hat{c}_{ijk}\sum_{u_{1},\ldots,u_{K}}^{K}~\sum_{\bar{u}_{1},\ldots,\bar{u}_{K}}^{K}~\sum_{\substack{i_1,\ldots,i_K\\\in\mathcal{I}[i,j,k]}}~\sum_{\substack{\bar{\imath}_1,\ldots,\bar{\imath}_K\\\in\mathcal{I}[i,j,k]}}\epsilon_{u_{1},\ldots,u_{K}}\epsilon_{\bar{u}_{1},\ldots,\bar{u}_{K}}\epsilon_{i_1,\ldots,i_K}\epsilon_{\bar{\imath}_1,\ldots,\bar{\imath}_K}\delta_{i_1,\bar{\imath}_1}\ldots\delta_{i_K,\bar{\imath}_K}\times\\[2mm]
&\hspace{6cm}(\tilde\nabla_{i_1}p_{u_1}\cdot\tilde\nabla_{\bar{\imath}_1}\bar{p}_{\bar{u}_1})\ldots(\tilde\nabla_{i_K}p_{u_K}\cdot\tilde\nabla_{\bar{\imath}_K}\bar{p}_{\bar{u}_K}) \;,
\end{split}
\end{align} 
where the scalar product $(\nabla_{i}p_{u}\cdot\nabla_{\bar{\imath}}\bar{p}_{\bar{u}})$ is the standard Euclidean scalar product of the two vectors. Note that, since the $\nabla_{ia}$ in fact only depend on $i$, we have only attached a label $i$ to $\Lambda_{ijk}$. All $\Lambda_{ijk}$ that cannot be constructed in this way (for example, since it is impossible to delete three affine coordinates $a_1$, $a_2$ and $a_3$ from a $\mathbb{P}^1$ or $\mathbb{P}^2$) are zero. Furthermore, there are two symmetry factors $c_{ijk}$ and $\hat{c}_{ijk}$ in the expression $\eqref{eq:AffineLambdaijkGeneral}$. The $c_{ijk}$ arise from symmetries in the indices $i,j,k$ and are given by
\begin{align}
c_{ijk}=\left\{
\begin{array}{ll}
\frac{1}{1!} &\qquad \text{if $\lambda_{ijk}\neq0$ and all indices are the same}\\[1mm]
\frac{1}{3} &\qquad \text{if $\lambda_{ijk}\neq0$ and two indices are the same}\\[1mm]
\frac{1}{3!} &\qquad \text{if $\lambda_{ijk}\neq0$ and all indices are distinct}
\end{array}
\right.\;.
\eqlabel{eq:cijkSymmetryFactors}
\end{align}
The  factors $\hat{c}_{ijk}$ arise from an over-counting of different factors in the sums: since $\tilde\nabla_{ia}p_{u}$ is the same for all $a$, we get some terms several times. To be more precise, we generically get each factor $K$ times. However, it sometimes happens that by leaving out the indices $i,j,k$ in the sums, some $\mathbb{P}^{n_i}$ do not enter at all\footnote{Note that this can only happen for $\mathbb{P}^1$, $\mathbb{P}^2$, $\mathbb{P}^3$, if one, two, or all three indices $i,j,k$ are the same, respectively.} and hence do not give rise to a symmetry factor. We find that
\begin{align}
\hat{c}_{ijk}=\frac{1}{(K+1-\rho)!}\;,\quad\text{where}\quad\rho=\text{number of different $\mathbb{P}^{n_{i}}$ factors entering in the sum.}
\end{align}

Looking at Eq.~\eqref{JJJ}, we note the following: $\Omega_0$ on the right-hand side is given by Eq.~\eqref{eq:Omega0Generic} involve derivatives of the defining polynomials in the denominator. In contrast, the K\"ahler forms $J$ on the left-hand side do not explicitly involve $p_u$. A CICY given by the zero locus of a set of polynomials, $p_{u}(z)=0$, is left invariant by a scaling of $p_{u}\rightarrow\lambda_u p_{u}$. However, from Eq.~\eqref{JJJ} it would then seem as if the left-hand side does not scale with $\lambda_u$ while the right-hand side does. This is resolved by the observation that expression~\eqref{eq:AffineLambdaijkGeneral} scales as $\Lambda_{ijk}\rightarrow\prod_{u=1}^K|\lambda_u|^2\Lambda_{ijk}$. At the same time, $|\text{det}(B_K[r,s,t])|^2\rightarrow\prod_{u=1}^K|\lambda_u|^2~|\text{det}(B_K[r,s,t])|^2$. Since $|\text{det}(B_K[r,s,t])|^2$ appears in the denominator in Eq.~\eqref{eq:Omega0Generic}, Eq.~\eqref{JJJ} is homogeneous of degree~0.

The expression for the $\Lambda_{ijk}$ in terms of homogeneous coordinates is
\begin{align}
\eqlabel{eq:LambdaijkHom}
\begin{split}
\Lambda_{ijk}=&\frac{c_{ijk}}{\pi^3}\left(\frac{1}{\sigma_i\sigma_j\sigma_k}\prod_{l=1}^m\frac{1}{\sigma_l}\right)\times\\&\hat{c}_{ijk}\sum_{u_{1},\ldots,u_{K}}^{K}~\sum_{\bar{u}_{1},\ldots,\bar{u}_{K}}^{K}~\sum_{\substack{i_1,\ldots,i_K\\\in\mathcal{I}[i,j,k]}}~\sum_{\substack{\bar{\imath}_1,\ldots,\bar{\imath}_K\\\in\mathcal{I}[i,j,k]}}\epsilon_{u_{1},\ldots,u_{K}}\epsilon_{\bar{u}_{1},\ldots,\bar{u}_{K}}\epsilon_{i_1,\ldots,i_K}\epsilon_{\bar{\imath}_1,\ldots,\bar{\imath}_K}\delta_{i_1,\bar{\imath}_1}\ldots\delta_{i_K,\bar{\imath}_K}\times\\[2mm]
&\hspace{6cm}(\nabla_{i_1}P_{u_1}\cdot\nabla_{\bar{\imath}_1}\bar{P}_{\bar{u}_1})\ldots(\nabla_{i_K}P_{u_K}\cdot\nabla_{\bar{\imath}_K}\bar{P}_{\bar{u}_K}) \;.
\end{split}
\end{align}
where we use the notation explained in Eqs.~\eqref{skdef} and \eqref{pAndP}. Here, like in the analogous result for co-dimension one CICYs, Eq.~\eqref{L1res}, $\nabla_i P_u$ denotes the standard gradient of the $u^{\text{th}}$ polynomial with respect to the coordinates $x_{iA}$ of the $i^{\text{th}}$ projective ambient space factor. 

It is instructive to compare Eq.~\eqref{eq:LambdaijkHom} with the co-dimension one result~\eqref{L1res} in more detail. We note that the division by $|\nabla_i P|^2|\nabla_j P|^2|\nabla_k P|^2$ effectively leads to the omission of these terms, making it the analog of the omissions encoded by the index sets $\mathcal{I}[i,j,k]$. For the co-dimension one case, the combinatorial factors $c_{ijk}$ in Eq.~\eqref{eq:LambdaijkHom} do indeed specialise to the factors of the same name in Eq.~\eqref{L1res}. When comparing the affine expression for the $\Lambda_{ijk}$ in Eqn.~\eqref{eq:AffineLambdaijkGeneral} with the expression of the tetra-quadric~\eqref{LTQ}, it seems as if there is an additional factor of $\kappa_i$ in the denominator of Eq.~\eqref{eq:AffineLambdaijkGeneral}. However, the scalar products $(\tilde\nabla_i\,p_u\cdot \tilde\nabla_{\bar\imath}\,\bar{p}_{\bar u})$ lead to an extra $\kappa_i$, such that the two expressions match exactly.

We still need to demonstrate that the affine and homogeneous formulae for $\Lambda_{ijk}$ in Eqs.~\eqref{eq:AffineLambdaijkGeneral} and~\eqref{eq:LambdaijkHom} are indeed equivalent. To do this we note that
\begin{align}
\sigma_i=\sum_{A} |x_{iA}|^2=|x_{i0}|^2 \left(1+\sum_a z_{ia}\right)=|x_{i0}|^2 \kappa_i\; ,
\eqlabel{eq:sigmaToKappa}
\end{align}
and
\begin{align}
\nabla_i P_{u}\cdot \nabla_{\bar{\imath}}\bar{P}_{\bar{u}}=\frac{\mathfrak{s}_{u} \bar{\mathfrak{s}}_{\bar{u}}}{x_{i0}\bar{x}_{\bar{\imath}0}}
\left[\sum_{a,b}(\delta_{a,b}+z_{ia}\bar{z}_{\bar{\imath}b})p_{u,a}\bar{p}_{\bar{u},b}\right]=\frac{\mathfrak{s}_{u} \bar{\mathfrak{s}}_{\bar{u}}}{x_{i0}\bar{x}_{\bar{\imath}0}} (\tilde\nabla_i p_u\cdot \tilde\nabla_{\bar{\imath}}\bar{p}_{\bar{u}})\;,\quad\text{with}\quad \mathfrak{s}_u=\prod_{i=1}^m x_{i0}^{q^i_{u}}\;.\eqlabel{eq:NablaPProduct}
\end{align}
The full expression~\eqref{eq:LambdaijkHom} involves a product of $K$ scalar products of the type~\eqref{eq:NablaPProduct}. Due to the factors $\epsilon_{u_1\ldots u_K}$, $\epsilon_{\bar{u}_1\ldots \bar{u}_K}$ and $\delta_{i_j,\bar\imath_{j}}$, the structure functions $\Lambda_{ijk}$ pick up a pre-factor
\begin{align*}
\prod_{i_j,\bar\imath_j}^m~\prod_{u,\bar u}^K\frac{\mathfrak{s}_{u} \bar{\mathfrak{s}}_{\bar{u}}}{x_{i_j0}\bar{x}_{\bar{\imath_j}0}}=\prod_{i_j}^m~\prod_{u,\bar u=1}^K\frac{\mathfrak{s}_{u} \bar{\mathfrak{s}}_{\bar{u}}}{|x_{i0}|^2}\;.
\end{align*}
This can be further simplified by using the Calabi-Yau condition
\begin{align}
\sum_{u=1}^K q^i_{u}=n_i+1\;,
\end{align}
such that we get 
\begin{align}
\eqlabel{eq:NablaHomToNablaAff}
(\nabla_{i_1}P_{u_1}\cdot\nabla_{\bar{\imath}_1}\bar{P}_{\bar{u}_1})\!\ldots\!(\nabla_{i_K}P_{u_K}\cdot\nabla_{\bar{\imath}_K}\bar{P}_{\bar{u}_K})\!=\!\frac{|x_{i_10}|^2\ldots |x_{i_K0}|^2}{\prod_{l=1}^m|x_{l0}^{n_l+1}|^2}(\tilde\nabla_{i_1}p_{u_1}\cdot\tilde\nabla_{\bar{\imath}_1}\bar{p}_{\bar{u}_1})\!\ldots\!(\tilde\nabla_{i_K}p_{u_K}\cdot\tilde\nabla_{\bar{\imath}_K}\bar{p}_{\bar{u}_K})\;.
\end{align}
By writing the numerator $|x_{i_10}|^2\ldots |x_{i_K0}|^2$ as a product over all $d=K+3$ indices $(ia)$ and dividing by the three that are left out, we find
\begin{align}
|x_{i_10}|^2\ldots |x_{i_K0}|^2=\frac{\prod_{l=1}^m |x_{l0}^{n_l}|^2}{|x_{i0}|^2|x_{j0}|^2|x_{k0}|^2}\;.
\end{align}
Finally, we can use this to rewrite the pre-factor in \eqref{eq:NablaHomToNablaAff} as
\begin{align}
\frac{|x_{i_10}|^2\ldots |x_{i_K0}|^2}{\prod_{l=1}^m|x_{l0}^{n_l+1}|^2}=\frac{1}{|x_{i0}|^2 |x_{j0}|^2|x_{k0}|^2}\frac{1}{\prod_{l=1}^m|x_{l0}|^2}\;.
\end{align}
According to Eq.~\eqref{eq:sigmaToKappa}, this is precisely the factor that converts the $\sigma_i$ appearing in the denominator of Eq.~\eqref{eq:LambdaijkHom} into the $\kappa_i$ appearing in the denominator of Eq.~\eqref{eq:AffineLambdaijkGeneral}.


\bigskip
\noindent
\providecommand{\href}[2]{#2}


\begingroup\begin{thebibliography}{10}

\bibitem{Candelas:1987kf}
P.~Candelas, A.~M. Dale, C.~A. Lutken, and R.~Schimmrigk ``{Complete
  Intersection Calabi-Yau Manifolds},'' {\em Nucl. Phys.} {\bf B298} (1988)
493.

\bibitem{Green:1986ck}
P.~Green and T.~Hubsch ``{Calabi-Yau Manifolds as Complete Intersections in
  Products of Complex Projective Spaces},'' {\em Commun. Math. Phys.} {\bf 109}
  (1987)
99.

\bibitem{Hubsch:1992nu}
T.~Hubsch {\em {Calabi-Yau manifolds: A Bestiary for physicists}}.
\newblock World Scientific Singapore
1994.
\newblock

\bibitem{Kreuzer:2000xy}
M.~Kreuzer and H.~Skarke ``{Complete classification of reflexive polyhedra in
  four-dimensions},'' {\em Adv. Theor. Math. Phys.} {\bf 4} (2002) 1209--1230
\href{http://www.arXiv.org/abs/hep-th/0002240}{[{\tt hep-th/0002240}]}.

\bibitem{Yau:1798aa}
S.-T. Yau ``Calabi{\textquoteright}s conjecture and some new results in
  algebraic geometry,'' {\em Proceedings of the National Academy of Sciences}
  {\bf 74} (1977) no.~5, 1798--1799
  \href{http://www.arXiv.org/abs/http://www.pnas.org/content/74/5/1798.full.pdf}{[{\tt
  http://www.pnas.org/content/74/5/1798.full.pdf}]}.

\bibitem{Donaldson:2005aa}
S.~K. {Donaldson} ``{Some numerical results in complex differential
  geometry},'' \href{http://www.arXiv.org/abs/math/0512625}{[{\tt
  math/0512625}]}.

\bibitem{Douglas:2006rr}
M.~R. Douglas, R.~L. Karp, S.~Lukic, and R.~Reinbacher ``{Numerical Calabi-Yau
  metrics},'' {\em J. Math. Phys.} {\bf 49} (2008) 032302
\href{http://www.arXiv.org/abs/hep-th/0612075}{[{\tt hep-th/0612075}]}.

\bibitem{Braun:2007sn}
V.~Braun, T.~Brelidze, M.~R. Douglas, and B.~A. Ovrut ``{Calabi-Yau Metrics for
  Quotients and Complete Intersections},'' {\em JHEP} {\bf 05} (2008) 080
\href{http://www.arXiv.org/abs/0712.3563}{[{\tt 0712.3563}]}.

\bibitem{Douglas:2015aga}
M.~R. Douglas ``{Calabi--Yau metrics and string compactification},'' {\em Nucl.
  Phys.} {\bf B898} (2015) 667--674
\href{http://www.arXiv.org/abs/1503.02899}{[{\tt 1503.02899}]}.

\bibitem{2005math......8428B}
R.~L. {Bryant} ``{Remarks on the geometry of almost complex 6-manifolds},''
  {\em ArXiv Mathematics e-prints} (Aug., 2005)
  \href{http://www.arXiv.org/abs/math/0508428}{[{\tt math/0508428}]}.

\bibitem{Fernandez:2008wa}
M.~Fernandez, S.~Ivanov, L.~Ugarte, and R.~Villacampa ``{Non-Kaehler Heterotic
  String Compactifications with non-zero fluxes and constant dilaton},'' {\em
  Commun. Math. Phys.} {\bf 288} (2009) 677--697
\href{http://www.arXiv.org/abs/0804.1648}{[{\tt 0804.1648}]}.

\bibitem{Grantcharov:2009qv}
G.~Grantcharov ``{Geometry of compact complex homogeneous spaces with vanishing
  first Chern class},'' {\em Adv. Math.} {\bf 226} (2011) 3136--3159
\href{http://www.arXiv.org/abs/0905.0040}{[{\tt 0905.0040}]}.

\bibitem{Fei:2014aca}
T.~Fei and S.-T. Yau ``{Invariant Solutions to the Strominger System on Complex
  Lie Groups and Their Quotients},'' {\em Commun. Math. Phys.} {\bf 338} (2015)
  no.~3, 1183--1195
\href{http://www.arXiv.org/abs/1407.7641}{[{\tt 1407.7641}]}.

\bibitem{Otal:2016bgn}
A.~Otal, L.~Ugarte, and R.~Villacampa ``{Invariant solutions to the Strominger
  system and the heterotic equations of motion},'' {\em Nucl. Phys.} {\bf B920}
  (2017) 442--474
\href{http://www.arXiv.org/abs/1604.02851}{[{\tt 1604.02851}]}.

\bibitem{Goldstein:2002pg}
E.~Goldstein and S.~Prokushkin ``{Geometric model for complex nonKahler
  manifolds with SU(3) structure},'' {\em Commun. Math. Phys.} {\bf 251} (2004)
  65--78
\href{http://www.arXiv.org/abs/hep-th/0212307}{[{\tt hep-th/0212307}]}.

\bibitem{Fu:2006vj}
J.-X. Fu and S.-T. Yau ``{The Theory of superstring with flux on non-Kahler
  manifolds and the complex Monge-Ampere equation},'' {\em J. Diff. Geom.} {\bf
  78} (2008) no.~3, 369--428
\href{http://www.arXiv.org/abs/hep-th/0604063}{[{\tt hep-th/0604063}]}.

\bibitem{Fei:2017ctw}
T.~Fei, Z.~Huang, and S.~Picard ``{A Construction of Infinitely Many Solutions
  to the Strominger System},''
\href{http://www.arXiv.org/abs/1703.10067}{[{\tt 1703.10067}]}.

\bibitem{Martelli:2010jx}
D.~Martelli and J.~Sparks ``{Non-Kahler heterotic rotations},'' {\em Adv.
  Theor. Math. Phys.} {\bf 15} (2011) no.~1, 131--174
\href{http://www.arXiv.org/abs/1010.4031}{[{\tt 1010.4031}]}.

\bibitem{Nilsson:1984bj}
B.~E.~W. Nilsson and C.~N. Pope ``{Hopf Fibration of Eleven-dimensional
  Supergravity},'' {\em Class. Quant. Grav.} {\bf 1} (1984)
499.

\bibitem{SOROKIN1985301}
D.~Sorokin, V.~Tkach, and D.~Volkov ``On the relationship between compactified
  vacua of d = 11 and d = 10 supergravities,'' {\em Physics Letters B} {\bf
  161} (1985) no.~4, 301 -- 306.

\bibitem{Tomasiello:2007eq}
A.~Tomasiello ``{New string vacua from twistor spaces},'' {\em Phys. Rev.} {\bf
  D78} (2008) 046007
\href{http://www.arXiv.org/abs/0712.1396}{[{\tt 0712.1396}]}.

\bibitem{Koerber:2008rx}
P.~Koerber, D.~L\"ust, and D.~Tsimpis ``{Type IIA AdS(4) compactifications on
  cosets, interpolations and domain walls},'' {\em JHEP} {\bf 07} (2008) 017
\href{http://www.arXiv.org/abs/0804.0614}{[{\tt 0804.0614}]}.

\bibitem{Andriot:2015sia}
D.~Andriot ``{New supersymmetric vacua on solvmanifolds},'' {\em JHEP} {\bf 02}
  (2016) 112
\href{http://www.arXiv.org/abs/1507.00014}{[{\tt 1507.00014}]}.

\bibitem{Larfors:2010wb}
M.~Larfors, D.~L\"ust, and D.~Tsimpis ``{Flux compactification on smooth,
  compact three-dimensional toric varieties},'' {\em JHEP} {\bf 07} (2010) 073
\href{http://www.arXiv.org/abs/1005.2194}{[{\tt 1005.2194}]}.

\bibitem{Larfors:2011zz}
M.~Larfors ``{Flux compactifications on toric varieties},'' {\em Fortsch.
  Phys.} {\bf 59} (2011)
730--733.

\bibitem{Terrisse:2017tbb}
R.~Terrisse and D.~Tsimpis ``{SU(3) structures on S$^{2}$ bundles over
  four-manifolds},'' {\em JHEP} {\bf 09} (2017) 133
\href{http://www.arXiv.org/abs/1707.04636}{[{\tt 1707.04636}]}.

\bibitem{Gukov:1999ya}
S.~Gukov, C.~Vafa, and E.~Witten ``{CFT's from Calabi-Yau four folds},'' {\em
  Nucl. Phys.} {\bf B584} (2000) 69--108
  \href{http://www.arXiv.org/abs/hep-th/9906070}{[{\tt hep-th/9906070}]}.
[Erratum: Nucl. Phys.B608,477(2001)].

\bibitem{Witten:1985bz}
E.~Witten ``{New Issues in Manifolds of SU(3) Holonomy},'' {\em Nucl. Phys.}
  {\bf B268} (1986)
79.

\bibitem{Witten:1986kg}
L.~Witten and E.~Witten ``{Large Radius Expansion of Superstring
  Compactifications},'' {\em Nucl. Phys.} {\bf B281} (1987)
109--126.

\bibitem{li2005}
J.~Li and S.-T. Yau ``The existence of supersymmetric string theory with
  torsion,'' {\em J. Differential Geom.} {\bf 70} (05, 2005) 143--181.

\bibitem{Andreas:2010cv}
B.~Andreas and M.~Garcia-Fernandez ``{Heterotic Non-Kahler Geometries via
  Polystable Bundles on Calabi-Yau Threefolds},'' {\em J. Geom. Phys.} {\bf 62}
  (2012) 183--188
\href{http://www.arXiv.org/abs/1011.6246}{[{\tt 1011.6246}]}.

\bibitem{Andreas:2010qh}
B.~Andreas and M.~Garcia-Fernandez ``{Solutions of the Strominger System via
  Stable Bundles on Calabi-Yau Threefolds},'' {\em Commun. Math. Phys.} {\bf
  315} (2012) 153--168
\href{http://www.arXiv.org/abs/1008.1018}{[{\tt 1008.1018}]}.

\bibitem{2000math.....10054H}
N.~{Hitchin} ``{The geometry of three-forms in six and seven dimensions},''
  {\em ArXiv Mathematics e-prints} (Oct., 2000)
  \href{http://www.arXiv.org/abs/math/0010054}{[{\tt math/0010054}]}.

\bibitem{2002math......2282C}
S.~{Chiossi} and S.~{Salamon} ``{The intrinsic torsion of SU(3) and G\_2
  structures},'' {\em ArXiv Mathematics e-prints} (Feb., 2002)
  \href{http://www.arXiv.org/abs/math/0202282}{[{\tt math/0202282}]}.

\bibitem{LopesCardoso:2002vpf}
G.~Lopes~Cardoso, G.~Curio, G.~Dall'Agata, D.~Lust, P.~Manousselis, and
  G.~Zoupanos ``{NonKahler string backgrounds and their five torsion
  classes},'' {\em Nucl. Phys.} {\bf B652} (2003) 5--34
\href{http://www.arXiv.org/abs/hep-th/0211118}{[{\tt hep-th/0211118}]}.

\bibitem{Grana:2005jc}
M.~Gra\~na ``{Flux compactifications in string theory: A Comprehensive
  review},'' {\em Phys.Rept.} {\bf 423} (2006) 91--158
\href{http://www.arXiv.org/abs/hep-th/0509003}{[{\tt hep-th/0509003}]}.

\bibitem{Hull:1986kz}
C.~M. Hull ``{Compactifications of the Heterotic Superstring},'' {\em Phys.
  Lett.} {\bf B178} (1986)
357--364.

\bibitem{Strominger:1986uh}
A.~Strominger ``{Superstrings with Torsion},'' {\em Nucl. Phys.} {\bf B274}
  (1986)
253.

\bibitem{Ivanov:2000fg}
S.~Ivanov and G.~Papadopoulos ``{A No go theorem for string warped
  compactifications},'' {\em Phys.Lett.} {\bf B497} (2001) 309--316
\href{http://www.arXiv.org/abs/hep-th/0008232}{[{\tt hep-th/0008232}]}.

\bibitem{Giddings:2001yu}
S.~B. Giddings, S.~Kachru, and J.~Polchinski ``{Hierarchies from fluxes in
  string compactifications},'' {\em Phys. Rev.} {\bf D66} (2002) 106006
\href{http://www.arXiv.org/abs/hep-th/0105097}{[{\tt hep-th/0105097}]}.

\bibitem{Becker:2003yv}
K.~Becker, M.~Becker, K.~Dasgupta, and P.~S. Green ``{Compactifications of
  heterotic theory on nonKahler complex manifolds. 1.},'' {\em JHEP} {\bf 04}
  (2003) 007
\href{http://www.arXiv.org/abs/hep-th/0301161}{[{\tt hep-th/0301161}]}.

\bibitem{Gauntlett:2003cy}
J.~P. Gauntlett, D.~Martelli, and D.~Waldram ``{Superstrings with intrinsic
  torsion},'' {\em Phys.Rev.} {\bf D69} (2004) 086002
\href{http://www.arXiv.org/abs/hep-th/0302158}{[{\tt hep-th/0302158}]}.

\bibitem{Behrndt:2004km}
K.~Behrndt and M.~Cvetic ``{General N = 1 supersymmetric flux vacua of
  (massive) type IIA string theory},'' {\em Phys. Rev. Lett.} {\bf 95} (2005)
  021601
\href{http://www.arXiv.org/abs/hep-th/0403049}{[{\tt hep-th/0403049}]}.

\bibitem{Grana:2004bg}
M.~Gra\~na, R.~Minasian, M.~Petrini, and A.~Tomasiello ``{Supersymmetric
  backgrounds from generalized Calabi-Yau manifolds},'' {\em JHEP} {\bf 0408}
  (2004) 046
\href{http://www.arXiv.org/abs/hep-th/0406137}{[{\tt hep-th/0406137}]}.

\bibitem{Lust:2004ig}
D.~L\"ust and D.~Tsimpis ``{Supersymmetric AdS(4) compactifications of IIA
  supergravity},'' {\em JHEP} {\bf 02} (2005) 027
\href{http://www.arXiv.org/abs/hep-th/0412250}{[{\tt hep-th/0412250}]}.

\bibitem{Behrndt:2005bv}
K.~Behrndt, M.~Cvetic, and P.~Gao ``{General type IIB fluxes with SU(3)
  structures},'' {\em Nucl. Phys.} {\bf B721} (2005) 287--308
\href{http://www.arXiv.org/abs/hep-th/0502154}{[{\tt hep-th/0502154}]}.

\bibitem{Larfors:2013zva}
M.~Larfors ``{Revisiting toric SU(3) structures},'' {\em Fortsch. Phys.} {\bf
  61} (2013) 1031--1055
\href{http://www.arXiv.org/abs/1309.2953}{[{\tt 1309.2953}]}.

\bibitem{Blumenhagen:2006ci}
R.~Blumenhagen, B.~Kors, D.~Lust, and S.~Stieberger ``{Four-dimensional String
  Compactifications with D-Branes, Orientifolds and Fluxes},'' {\em Phys.
  Rept.} {\bf 445} (2007) 1--193
\href{http://www.arXiv.org/abs/hep-th/0610327}{[{\tt hep-th/0610327}]}.

\bibitem{Denef:2008wq}
F.~Denef ``{Les Houches Lectures on Constructing String Vacua},'' {\em Les
  Houches} {\bf 87} (2008) 483--610
\href{http://www.arXiv.org/abs/0803.1194}{[{\tt 0803.1194}]}.

\bibitem{Gauntlett:2002sc}
J.~P. Gauntlett, D.~Martelli, S.~Pakis, and D.~Waldram ``{G structures and
  wrapped NS5-branes},'' {\em Commun. Math. Phys.} {\bf 247} (2004) 421--445
\href{http://www.arXiv.org/abs/hep-th/0205050}{[{\tt hep-th/0205050}]}.

\bibitem{delaOssa:2014cia}
X.~de~la Ossa and E.~E. Svanes ``{Holomorphic Bundles and the Moduli Space of
  N=1 Supersymmetric Heterotic Compactifications},'' {\em JHEP} {\bf 10} (2014)
  123
\href{http://www.arXiv.org/abs/1402.1725}{[{\tt 1402.1725}]}.

\bibitem{Witten:1985xc}
E.~Witten ``{Symmetry Breaking Patterns in Superstring Models},'' {\em Nucl.
  Phys.} {\bf B258} (1985)
75.

\bibitem{Strominger:1985it}
A.~Strominger and E.~Witten ``{New Manifolds for Superstring
  Compactification},'' {\em Commun. Math. Phys.} {\bf 101} (1985)
341.

\bibitem{Anderson:2017aux}
L.~B. Anderson, X.~Gao, J.~Gray, and S.-J. Lee ``{Fibrations in CICY
  Threefolds},'' {\em JHEP} {\bf 10} (2017) 077
\href{http://www.arXiv.org/abs/1708.07907}{[{\tt 1708.07907}]}.

\bibitem{Kachru:2002sk}
S.~Kachru, M.~B. Schulz, P.~K. Tripathy, and S.~P. Trivedi ``{New
  supersymmetric string compactifications},'' {\em JHEP} {\bf 03} (2003) 061
\href{http://www.arXiv.org/abs/hep-th/0211182}{[{\tt hep-th/0211182}]}.

\bibitem{Schulz:2004ub}
M.~B. Schulz ``{Superstring orientifolds with torsion: O5 orientifolds of torus
  fibrations and their massless spectra},'' {\em Fortsch. Phys.} {\bf 52}
  (2004) 963--1040
\href{http://www.arXiv.org/abs/hep-th/0406001}{[{\tt hep-th/0406001}]}.

\bibitem{Caviezel:2008ik}
C.~Caviezel, P.~Koerber, S.~Kors, D.~L\"ust, D.~Tsimpis, and M.~Zagermann
  ``{The Effective theory of type IIA AdS(4) compactifications on nilmanifolds
  and cosets},'' {\em Class. Quant. Grav.} {\bf 26} (2009) 025014
\href{http://www.arXiv.org/abs/0806.3458}{[{\tt 0806.3458}]}.

\bibitem{McOrist:2012yc}
J.~McOrist and S.~Sethi ``{M-theory and Type IIA Flux Compactifications},''
  {\em JHEP} {\bf 12} (2012) 122
\href{http://www.arXiv.org/abs/1208.0261}{[{\tt 1208.0261}]}.

\bibitem{Petrini:2013ika}
M.~Petrini, G.~Solard, and T.~Van~Riet ``{AdS vacua with scale separation from
  IIB supergravity},'' {\em JHEP} {\bf 11} (2013) 010
\href{http://www.arXiv.org/abs/1308.1265}{[{\tt 1308.1265}]}.

\bibitem{Avez:70}
A.~Avez ``Characteristic {C}lasses and {W}eyl {T}ensor: {A}pplications to
  {G}eneral {R}elativity,'' {\em Proceedings of the National Academy of
  Sciences of the United States of America} {\bf 66} (1970) no.~2, 265--268.

\end{thebibliography}\endgroup
\end{document}